# Charles Michie Smith – founder of the Kodaikanal (Solar Physics) Observatory and beginnings of physical astronomy in India*

*N. Kameswara Rao, A. Vagiswari and Christina Birdie*

*It has been about 115 years since the establishment of Kodaikanal Observatory as an extension of the original Madras Observatory, which has evolved into the present Indian Institute of Astrophysics at Bangalore. It is also the first mountain observatory in India. Charles Michie Smith was the man who selected the site, established the observatory and directed it for the first 12 years. He was also the man who recruited John Evershed, discoverer of the famous Evershed effect and established Kodaikanal Observatory as a major centre for solar physics. Michie Smith, the person and the establishment of Kodaikanal Observatory are discussed here in the context of the early studies in physical astronomy (observational astrophysics) in India.*

'Solar Physics was born in the tobacco fields of Guntur' declared Vainu Bappu[1], reminiscing about the famous discovery of the element helium during the total solar eclipse of 18 August 1868, that passed through Guntur in Andhra Pradesh, India. A spectroscope was used for the first time at a total solar eclipse to look at the limb prominences during an eclipse. The prominence spectra showed the presence of an emission line that did not correspond to any known element on Earth till that time. Both Janssen[2,3] (note 1) (who made spectroscopic observations at Guntur during the eclipse) and Norman Lockyer (who made spectroscopic observations of prominences outside the eclipse) were credited as the discoverers of helium[4,5]. Soon after the eclipse, astonished by the brilliance of the red line of hydrogen in the prominences during the eclipse, Janssen looked for it in full sunshine through a spectroscope and found it.

The Madras Observatory also established an eclipse camp at Masulipatam where Norman Pogson, the famous Madras astronomer, made spectroscopic observations and discovered the emission line (D3) in the prominence spectrum. He sketched the spectrum and noted that the emission line did not coincide with the famous D lines in the solar spectrum (Figure 1). The visibility of red prominences (Figure 2) after reappearance of the Sun was also confirmed. Little known, but a most important fact about the experiments at Pogson's Muslipatam camp is 'the detection of bright lines in the solar corona (Figure 3), especially of the most important one now known as Kirchoff 1474 (note 2, ref. 6). The bright lines were seen simultaneously by other astronomers, but actual micrometrical measurements of their positions were obtained with spectroscope at Masulipatam'[7] Had it not been for the red-tape methods of conducting Government publications (at that time), he (Pogson) would have got the credit of 'being the first to detect the bright lines of the spectrum of corona at the Indian eclipse of 1868' (ref. 8). The discovery of this coronal line has been credited to C. A. Young in the total solar eclipse of 7 August 1869.

However, systematic observational studies of solar physics could be pursued in the country only after the establishment of the Kodaikanal Observatory in 1899 (note 3, ref. 9) (Solar Physics). The man who selected the site, established and equipped the observatory in the face of major difficulties was Charles Michie Smith, a professor of physics at Madras Christian College, who slowly got lured into astronomy and became the Government astronomer at Madras Observatory first and then the founder Director of

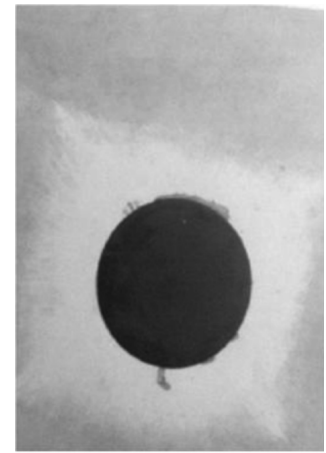

**Figure 2.** Prominences at the total solar eclipse of 18 August 1868 as observed by Pogson at Masulipatnam camp of Madras Observatory. Pogson painted this picture.

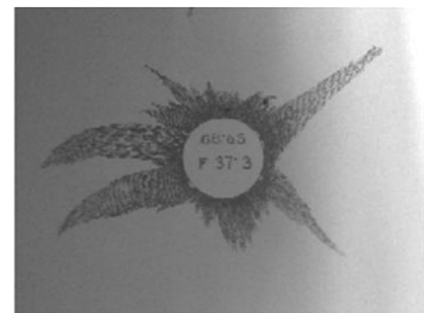

**Figure 3.** Drawing of the corona at the time of total solar eclipse of 18 August 1868 – from Naegamvala[11].

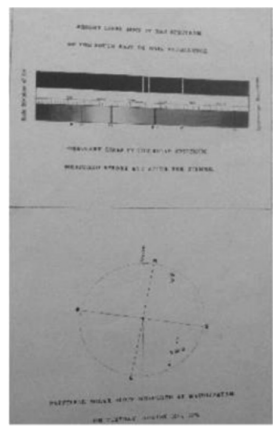

**Figure 1.** Spectra of the prominence at the total solar eclipse of 18 August 1868 as observed by Pogson at Masulipatnam camp of Madras Observatory. Pogson painted the picture (spectrum).

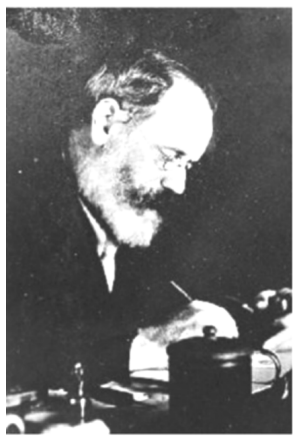

**Figure 4.** Charles Michie Smith.

Kodaikanal Observatory later. (Even after retirement from the observatory he continued to live in Kodaikanal where he died and was buried there.) According to Chinnici[10] '...establishment of Kodaikanal solar observatory, which was built in 1895 and started functioning in 1900, thanks to Charles Michie Smith (1854–1922) (Figure 4), Director of Madras Observatory, who tried to start the practice of "New Astronomy" in India.'

## Physical astronomy (astronomical spectroscopy) in India before Kodaikanal

After the trigger of 1868 total solar eclipse, the emergence of new astronomy (spectroscopic astrophysics) in the country was essentially limited to special events like eclipses or transits of planets, etc. mostly by visiting foreign astronomers, sometimes in collaboration with resident astronomers, done in sporadic fashion at different places in the country, e.g. during the transit of Venus in 1874.

### Madras Observatory

The earliest recorded attempt at observations of stellar spectra was apparently in 1871, on the eve of the total solar eclipse, at Madras by (Lorenzo) Respighi and Pogson of the star $\gamma$ Argus (now Velorum). 'The discovery of the very remarkable nature of the bright star $\mu(\gamma)$ Argus (note 4, ref. 11) by Professor Respighi with a direct vision spectroscope applied to the large equatorial, must not be omitted. The spectrum of this star, unlike most others, consists of bright lines only, three being most exceptionally vivid, two in the red, and one in the blue part of the spectrum. Several other faint lines are clearly discernible, but subsequent measurements has shown that the deep red lines in the spectrum of this star by no means correspond with the red portion of the solar spectrum, but occupy instead the place of the yellow in the latter, while the vivid blue line is far beyond the well known hydrogen line F in Kirchoff's chart, almost approaching to the violet end of the solar spectrum. A star in Corona Borealis, which suddenly blazed out in 1860 (note 5) and faded away again in a few weeks, excited intense astonishment among European astronomers and physicists, and proved to owe its brief brilliancy to the temporary combustion of hydrogen gas; in fact to be a world on fire *for a short time*; but in $\mu(\gamma)$ Argus we have before us a *permanent furnace world*, the heat of which probably surpasses all conception, and is maintained by burning gases of a totally different nature to any with which we are acquainted.'[12] It is now believed, that the combustion referred by Pogson in T CrB is nuclear combustion rather than a chemical one as Pogson implied, but still, an energy generation from hydrogen in explosive way!

Subsequently, a total and an annular eclipse of the Sun, that occurred on 12 December 1871 and 6 June 1872 respectively, which passed through South India were also followed with a spectroscope as well as direct photography. Madras Observatory attempted for the first time celestial and solar photography during the 12 December 1871 eclipse. Pogson's son, Norman Everard Pogson, assistant to the Government astronomer, was especially sent to England to get trained with Warren De la Rue in solar photography that is to be carried out at this eclipse as well as coming events, like the next annular eclipse and 1874 transit of Venus. Three fair photographs were obtained during totality with nine-inch silver glass reflecting telescope by Browning. 'The photographs verify the existence of a luminous envelope around the sun's disc, to the height of about 160,000 miles and show other interesting features as well.'[12] However, the spectroscopic observations were not so successful, 'the consequence was, that of the bright lines seen in the spectroscope no measures were taken. One vivid but not very broad line, somewhere in the green part of the spectrum, was visible right across the field when the Sun's limb was clear of the slit. This was probably the Coronal line known as K1474, and was the only one which did not seem to change: the rest were flickering in and out, sometimes only three and sometimes seven or more, as the telescope was made to approach the moon's limb. These shorter and less steady lines were most likely prominence lines only, as the slit must have passed over several in sweeping about and around the limb.' It is sad that none of this work of the Madras Observatory has seen the light of the day. Pogson comments in 1874 May, 'It is a matter of great regret to me that continued pressure of work and distressing circumstances beyond all human control have, up to present time, prevented me from submitting the report upon this expedition.' C. Ragoonatha Charry[13] was a key member at this eclipse expedition and observations of 'ordinary telescope phenomena' were entrusted to him.

Several expeditions and observers conducted experiments during this eclipse from different places in South India. While Madras observers were at Avenashy, in Coimbatore district, Norman Lockyer set up camp at Baicull in South Canara and Tennant and others were at Doddabetta on the Neilgherry Hills, some were at Poodoocottah. One of the 'most important' observations during the 1871 eclipse was the detection, for the first time, of Fraünhofer lines (in absorption) in the coronal spectrum by Janssen located in the Nilgiris, besides the bright lines that were known to be present. This observation suggested that besides hot gas there is something that scatters the solar photospheric spectrum in the corona (later designated as K-Corona). In a way the polarization observations of Madras Observatory by G. K. Winters during this eclipse suggested the same 'increasing of polarization at a distance from sun's limb or in other words, strong reflection of solar rays from the higher and cooler regions of the Corona than from heated luminous shell shown in the photographs'[12]. Both Lockyer and Respighi introduced new instrumentation at this eclipse, namely slitless spectroscopy which produced the spectrum instead of a series of bright lines, produced a series of coloured rings, as the corona round the Sun is seen in each line. This way not only the whole corona is seen at once, but differences of height of the corona in different lines could also be observed[3].

Like Pogson, both (Lieutenant) John Herschel Jr and (Major) Tennant who were stationed at Doddabetta also noted that the coronal line 1474K could be seen all over the corona independent of the structure.

'The Annular Eclipse of the Sun on 6th June (1872) yielded results equal if not superior in interest to those of the total Eclipse expeditions of 1871. Four excellent photographs were taken about the annular phase, one of which shows the moon's dark limb rendered visible by the light of solar corona, before the completion of the ring. The operators were the late N. E. Pogson and F. Doderet. The phenomenon known as *reversal of the lines* hitherto observed only at total eclipses, was also seen to great advantage with the spectroscope at both formation and braking of the ring.'[14]

Even after these trials of eclipse spectroscopy, continuity in observations of either solar or stellar (celestial) spectroscopy was not maintained at Madras Observatory or elsewhere in the country until another special event occurred, namely the transit of Venus in 1874. However, occasionally, sunspots or some unusual solar features used to attract the attention of Madras astronomers. Such an event occurred on 17 April 1882. 'A most interesting solar spot, of vast size and subject to striking changes while under actual observation, was first seen on April 17th; attention being drawn to it by corresponding magnetic disturbance of considerable extent. Careful drawings of this spot were made on April 17th, 18th, 19th, 21st and 22nd, by Miss E. Isis Pogson, Mr A. L. Pogson and Mr R. F. Chisholm, the Government Architect.'[15]

In later years even this occasional activity seemed to have ceased and has been criticized severely. According to *Indian Engineering*[16] dated 18 February 1899 'For nearly ten years, the Solar Physics Committee, the secretary of state, and the Government of India have been under the impression that Solar Spectroscopic investigations were under taken at Madras. Nothing of that sort has been done'.

*Calcutta Observatory – The Italian connection*

The nineteenth century's first transit of Venus on 6 December 1874 was visible in most parts of India. It not only aroused considerable interest as an astronomical event for measuring the solar distance (astronomical unit), but also led to the establishment of Calcutta Observatory where solar spectroscopy (physics) was to be studied regularly. A team of astronomers led by Pietro Tacchini, from Italy, planned to follow the transit of Venus through a spectroscope as it passed over the Sun's disk and sought a suitable place for observations in India (see Chinnici[10] for a full account). He contacted various Italian Consuls at Bombay, Calcutta, Madras as well as Norman Pogson, then Director of Madras Observatory. The Consul at Calcutta, F. Lamouroux, responded first after consulting father Eugene Lafont, the famous Rector and Professor of Physics at Xavier's College in Calcutta. They suggested Madhupur (Muddapur) near Calcutta as a possible site. Lafont was a great promoter and profounder of science who co-founded, in 1876, the Indian Association for the Cultivation of Science and was rightly considered as one of the architects of modern Indian science[17]. He became part of the Italian observing team consisting of Jesuit Angelo Secchi (1818–78), Director of Collegio Romano (who later had to leave the expedition for health reasons), Alessandro Dorna (1825–86) of the Observatory of Turin, and Antonio Abetti (1846–1928) of the Observatory of Padua. During the transit, visual observations through a telescope were done by Dorna and Lafont, whereas Tacchini and Abetti carried out spectroscopic observations simultaneously. The main result was the difference in contact times recorded visually by a simple telescope and those recorded by the spectroscope. The spectroscopic ones were earlier by 2 min for the third and fourth contacts.

Tacchini also observed absorption bands in Venus spectrum during the transit that suggested existence of an atmosphere for Venus consisting of 'a great deal of vapours, analogous to the terrestrial ones'. Lafont gave a detailed account of it and concluded that the expedition showed 'upon experimental proofs, the great superiority of the spectroscopic method over all others, in determining the real time of contact, to a small fraction of a second, with ease and certainty. The main object of this mission is therefore accomplished'. Madras observatory's efforts to observe the transit were not successful because of bad weather. 'Venus was briefly seen once or twice during the transit, but only through thick clouds which rendered photographs or measurement of any kind impossible. The second internal contact noted by Miss E. Isis Pogson and C. Ragoonatha Charry was the only record obtainable after all the trouble incurred.'[14]

Tacchini remained in India even after the Venus transit and persuaded Lafont to build an observatory for monitoring the Sun, its spots and other features on and at the limb visually (by producing drawings) as well as spectroscopically on a regular basis to complement the observations that are already being done at Palermo and other Italian observatories. It would have been a collaborative effort and an extension of the Italian experiments at a different longitude providing continuity and complementarity to the observations (at Calcutta, solar phenomena would be visible about five hours earlier). In fact, Lafont did manage to build the observatory. Writing to Tacchini in July 1875 he remarked 'I am pleased to announce to you that our observatory is almost completed... Mr Merz had already written to me and he is busy building a 7-inch equatorial with parallactic mounting for 12,500 francs. It will not be finished before eighteen months. I am going to receive a 10 prism spectroscope of Browning as the Lockyer's one, and also a direct vision spectroscope, which I will use with a small 3-inch telescope of Steinheil, while waiting for the installation of my grand instrument.' Establishment of this spectroscopic observatory was announced by Tacchini 'The eminent father Lafont, Director of St Xavier's College in Calcutta, after observing the chromosphere and solar prominences with our instruments in Madhupur and seeing the practical way to execute the spectroscopic observations of the Sun at our station, has accepted the proposal to build an Observatory in Calcutta in his college with the aim of carrying out other regular solar observations, which … could fill the inevitable gaps of our Observatories… the new Calcutta observatory will be able to give the best results under the active direction of Lafont, to whom our colleagues will be very grateful for remedying, with his ability and commitment, a long complained snag.'[10] In anticipation of the arrival of the telescope, Lafont even started practising spectroscopic observations of prominences with a direct vision spectroscope attached to a 3-inch telescope. The establishment of the spectroscopic observatory was also announced in an article in *Nature* by Raphael Meldola in 1875. However, things got delayed and first results could only be obtained in 1878, which Lafont sent to Tacchini with a note not to publish them as they are his first trials. Based on these results Tacchini wrote, 'In this first essay sent by Lafont there are some solar limbs drawn in November 1877 and April 1878. The traits of the prominences and chromosphere were in good accordance to the drawings made in Palermo. ... drawings which we shall receive from Calcutta are entirely comparable with those made in Italy...'.

Those observations were never published and before any spectroscopic observations could begin in a regular manner, Lafont was taken ill and had to go to Europe. After 1878, for various reasons, the Italian connection became weak and no real spectroscopic observations ever came out of Calcutta Observatory. Later, Lafont's interests got diversified and he passed on the directorship of the observatory to others at St. Xavier's College. Writing a foreword to a publication of St. Xavier's College's Solar Observatory in December 1881, Lafont states '...but a severe illness caused by over-exposure to the Sun prevented the undersigned (Lafont) from continuing his labours and obliged him to give over to some of his colleagues. Rev. C. De Clippeleir, S. J. and Rev. V. De Campignelles, S. J., have now began a systematic record of the Solar spots and protuberances, which they intend publishing in the form of diagrams showing the positions, number, and principal details of these interesting features of our luminary. It has been deemed unnecessary to draw more than general outline of the protuberances (Figure 5), as their forms change with such constant rapidity, that a record of their fitting shapes cannot be of much scientific value. It is hoped that these humble contributions to the knowledge of Solar Physics may prove of some assistance towards the formation of a theory and help in completing the work of other Observatories, where the Sun is a special object of study'. The records presented are for December 1881–April 1882 by C. DeClippeleir, the Director.

Alphonse De Penaranda and Constain De Clippeleir took charge of the observatory during 1878–79 and 1882 onwards respectively. Nothing substantial emerged from the observatory even though it had good set of instruments available, except few reports of solar eclipses; annular eclipse of 17 July 1890 observed at Bhagalpur and the total eclipse of 22 January 1898 at Dumraon in Bihar. A brief description of this eclipse by the team of Jesuit priests, including those from St. Xavier's College is given in Biswas[17]. Occasional burst of activity took place triggered by some dramatic astronomical events like 'Gigantic solar disturbances were observed at Madras Observatory yesterday (25 April 1882): at the St. Xavier's college observatory something better was done – the disturbances were recorded by electric pen.'[17]. The observatory's activities slowly faded away. As remarked by Lafont in his letter to Tacchini '...I think that misfortune is attached to my observatory...' just about sums up the status of the observatory. Very few spectroscopic observations were carried out in spite of the fact that it was 'the only spectroscopic laboratory in India'[18], until Naegamvala established Takhatasingji Observatory in 1888 and started making spectroscopic observations of sun and stars.

*Takhatasingji Observatory – Poona*

Kavasji Dadabhai Naegamvala[19] (note 6) was a pioneering astrophysicist from Elphinston College, Bombay (Figure 6), who got inspired to pursue astronomical spectroscopy after visiting Lafont and his observatory in Calcutta in 1882/83. By 1888, he established the Maharaja Takhatasingji Observatory in Poona with the 'most modern equipment at that time in India' for astronomical spectroscopy and photography, and could legitimately be called India's first astrophysical observatory. 'The principal instruments are a 16½ inch silver-on-glass Newtonian by Sir H. Grubb, with a 4-inch finder attached and a 6-inch equatorial refractor by elder Cooke. Both these instruments are of highest excellence, and besides eyepieces and micrometers, they are furnished with several spectroscopes by Grubb, Hilger, Browning and others... It is intended at present to restrict the work of the observatory to certain branches of spectroscopic research, together with occasional observations of comets & c'[20].

Naegamvala got himself familiarized with spectroscopic equipment and methods of observation of the Sun and stars through his visits abroad and working with accomplished people like Norman Lockyer, Vogel and William Huggins, the three leading spectroscopists at that time. He was in frequent communication with other astronomers, notably then Astronomer Royal W. H. M. Christie. His spectroscopic observations of Orion nebula[21] showed that the green nebular line (note 7) is sharp, symmetrical and narrow and not 'fluted' under examination with several spectroscopes and magnifications with the 16½ inch telescope, thus putting the 'meteoritic hypothesis' of the nebulae by Norman Lockyer to the sword. According to this hypothesis, the lines arise from collisional heat of the meteoric particles, and require the nebular line to be fluted (extended or shaded like molecular bands – the likely molecular band at this wavelength is that of MgO). Even before Naegamvala's observations, others notably Huggins, Vogel and James Keeler at Lick showed the lines to be sharp, however there still was a major controversy. As Osterbrock[22] narrates 'although Keeler's paper convinced most of his contemporaries, Lockyer and his partisans could not accept the result, and at a meeting of the Royal Astronomical Society on 8 May 1891, with neither Huggins nor Lockyer present, an argument welled up. It began with the reading of a paper from K. D. Naegamvala of Poonah, India, who had been observing the Orion nebula with his 16½-inch telescope and three-prism spectroscope and found that the chief nebular line was sharp under all circumstances, and therefore not the remnant of magnesium fluting, as Lockyer had suggested. Captain William Noble, a friend and partisan of Huggins, rose and smoothly congratulated Naegamvala, through the Secretary who had read the paper.' Noble is supposed to have said 'The theory (meteoric)... has been already three times killed by Dr Huggins in England, Dr Vogel in Germany and Professor Keeler at Lick observatory, and I think that we must look on Mr Naegamvala as having finally killed and buried it' (see Osterbrock[22], for a more detailed account).

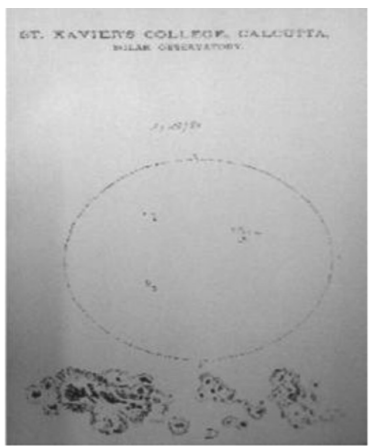

**Figure 5.** Contours of the sunspots recorded at Calcutta Observatory on 5 December 1881 by C. DeClippeleir and V. De Campignelles.

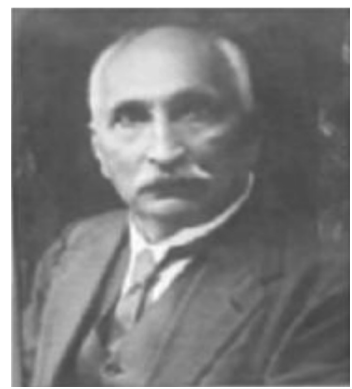

**Figure 6.** K. D. Naegamvala (photograph from a portrait of Naegamvala that was available with his grandson and daughter-in-law, obtained with their permission.)

Occasionally Naegamvala also observed the Sun and its spots spectroscopically. He observed emission reversals in the lines of hydrogen and D3 in absorption around the spots (great sunspot groups of February 1892) probably taken place during a flare build-up[23]. This was accompanied by the 'largest magnetic storm in recent years'. He also observed transit of Mercury on 9 May 1891. However, no systematic observational programme was carried out at the observatory. His most impressive and important contribution was the observation of the total solar eclipse of 22 January 1898 (see later).

In a letter written on 10 October 1896 (note 8), Naegamvala mentions 'By April next the equipment of the observatory will consist of (1) a 20″ Reflector 11′–3″ focus, with a six-inch guider, mounted on a stand very nearly the same as that of Grubbs 13″ astro-photographic telescopes.' In an application dated 10 January 1901, he proposes to observe with 'twenty-inch Cassegrain Equatorial of the observatory', which suggests that a replacement of the 16½ inch mirror by a 20 inch common mirror was in progress[24,25] (note 9).

In a letter addressed to W. H. M. Christie, the Astronomer Royal, dated 25 April 1895, Naegamvala emphasizes, 'I may here mention that this Observatory is the only establishment of its kind in India, where Photographic and Spectroscopic work could be carried out'. He makes his intentions clear with a follow-up letter dated 10 October 1896: 'My earnest desire is that this splendid equipment that I have managed to bring together should not lie idle', further as seen in the letter of 1901 'I am now relieved from instructional duties. I am now in a position to devote all my time to astronomical work... in case I am so fortunate as to merit its (committee's) confidence and support, I will to the best of my ability endeavour to employ the instrument for the furtherance of stellar spectroscopic research for which the climatic conditions of this place are so very favourable'. However, this promise was not kept up either because Naegamvala's work load as a teacher and as Director of the observatory was so much that he was unable to keep up the research work at the observatory or his interests waned with time. He had no collaborator or assistant to share research interests and keep up the continuity. Except for an occasional paper now and then, like the one describing the interesting spectrum of the Nova Persei 1901 (= GK Per) he obtained showing P-Cygni type spectral lines[26,27] (note 10), the activities of the observatory remained considerably low. After Naegamvala's retirement in 1912, the observatory was dismantled and the telescope was shifted to Kodaikanal.

### Other stations

In 1869 J. B. N. Hennessey of the Trigonometric Survey of India, commenced a series of actinometrical and spectroscopic investigations of the Sun at the instance of the members of Solar Physics Committee (see later) at Dehra Dun. Observations of measuring the solar radiation at the Earth and photography of the Sun's disc (features) on every clear day were routinely made at Simla and at Hennessy Observatory, Dehra Dun respectively. Photography of the Sun's surface had commenced from May 1878 under the supervision of Hennessey, at first with a small photoheliograph giving a 4-inch solar image. Later in 1882, bigger images were obtained with larger photoheliographs giving an 8-inch size disc on 12-inch plates[28].

The spectroscopic investigations were mainly concerned with the red end of the solar spectrum: beyond the Fraünhofer's A band. A series of papers in *Proceedings of the Royal Society* from 1871 to 1876, describe these efforts, mainly concerning how to discern atmospheric lines in the solar spectrum. His observations were conducted from 'a site in Himalayas at an altitude of 7100 feet (Dehra Dun) with a three-prism compound spectroscope specially constructed for him with the support of Royal Society by the Dublin firm of Grubb'[29]. Apparently, Piazzi Smyth was 'disappointed by Hennessey's spectral map published in *Transactions of the Royal Society*' and went ahead to Lisbon to observe the Sun at better resolving power in drier climate with very little water vapour.

### Background for solar monitoring and solar–terrestrial relations in the Indian context

A proposal to start an astrophysical observatory (physical astronomy) in India was being discussed in scientific circles in Briton since 1872 (Huggin's proposal to the Royal Astronomical Society[3]). Astronomical physics in the specific guise of solar physics was thought to be important in the context of possible relationship of solar phenomena to meteorological conditions (rainfall, etc.).

In 1876, a large deputation from the British Association had urged the Committee of Council on Education to carry out relevant recommendations of Devonshire Commission, including establishment of a physical observatory (Prime Minister Gladstone's ministry set up, in 1870, a Royal Commission on Scientific Instruction and the Advancement of Science under the 7th Duke of Devonshire (Chairman) with eight other members, including W. A. Miller (then treasurer of the Royal Society) later replaced by Smith, Stokes, Sharpley (secretaries of Royal Society), Huxley, and Lockyer). In the following year (1877) several leading scientists (including J. C. Adams, Joule, Maxwell, Roscoe, Stewart and Sir William Thomson) submitted a memorandum to the Committee of Council, in which the importance to meteorology of a solar physics observatory was stressed and a need for early decision was pleaded. The Committee then passed on the memorandum to Stokes, Balfour Stewart and General Strachey for their opinion.

Lockyer had already expressed himself as strongly in favour of having regular observations of the Sun made from somewhere in India. Stokes and Balfour Stewart's report clearly supported him. They suggested that astronomical physics should be divided into two areas. The first involved experimental research and the perfecting of observational methods; the second consisted rather continuous recording of solar phenomena. Stokes and Stewart emphasized that continuous solar observations were best made from North India. They noted that already 'arrangements have been made for sending out to India a highly intelligent sapper of the name Meins, who has been trained by Mr Lockyer, and will be employed in taking photographs of the Sun'[3]. Lockyer had obtained this concession from Lord Salisbury, who was Secretary for India in Disraeli's second administration. According to *Indian Engineering*[16] dated 18 February 1899 'in 1877, Lord Salisbury, then secretary of state for India asked Sir Norman Lockyer to formulate a programme of solar investigations to be undertaken in India'. Salisbury was sympathetic to the claims of science and had accepted the case for solar photography on a continuing basis in India. Subsequently, sapper Meins has been dispatched to India where, under the direction of the Surveyor-General, he began to take daily photographs of the Sun (in Simla). These were subsequently dispatched back to South Kensington for examination.

During 1878, an important move forward was made. The Duke of Devonshire appealed again to the Committee of Council on Education, asking them specifically to provide funds for the development of astronomical physics. A response was evoked at last. A Solar Physics Committee was established and charged with the duties of trying out new methods of observation, of determining what solar researches were underway in other countries, and of collecting and reducing observations of the Sun, especially those being made by Meins in India. The Committee consisted of six members, including the three advisers of the Committee of Council on Education: Stokes, Stewart, Strachey, and three representatives of the Science and Art Departments, one of whom was Lockyer, and one representative from Greenwich.

Attempts to correlate solar cycle with meteorological and geophysical parameters have been going on for some time. Piazzi Smyth published results of a long series of measurements, that started in 1837, of temperature just below the Earth's surface, using thermometers buried in the rock at Edinburgh. He concluded that there was a long-period variation in these temperatures with the same period as the solar cycle. Baxendell at the Radcliffe Observatory, Oxford, published an analysis of the meteorological records there for the previous 11 years, and claimed that there were variations of both atmospheric pressure and temperature in step with solar activity. At about the same time Stone, at the Cape of Good Hope, examined observations of the previous 30 years there and concluded that the mean annual temperature at the Cape was related to the solar cycle. In 1872, Meldrum in Mauritius used observation of many years there to show that cyclones in the Indian Ocean occurred most frequently when the sunspot number was largest, and that the rainfall in Mauritius, Adelaide and Brisbane was generally greater at sunspot maximum than at minimum. Lockyer used the rainfall records from the Cape and Madras and a variation similar to Meldrum. The importance of this result, if it could be established, is obvious. The years of minimum rainfall in India naturally also tended to be famine years. If a cycle of rainfall could be established, planning for future famines could be possible. Stewart also arrived at a result from river depths that maximum rainfall occurred soon after sunspot maximum and minimum. Stewart was also interested in the measurement of the solar heat incident on the Earth (the solar constant) in the hope of detecting variations which might lead to corresponding changes in climatic conditions. This work known as 'actinometry' was thought to be relevant by Lockyer and others to the development of solar physics in India, as the country provides clear skies required for such measurements.

Establishing a connection between meteorology and solar physics is thought to be not only of scientific importance, but also for economic reasons which were important for countries like India. In 1879 Lockyer and his colleagues presented to the Indian Famine Commission a report dealing with the relationship between the sunspot cycle and rainfall in South India. This report pointed out that famines in Madras since 1810 could be correlated closely with sunspot minimum. The authors emphasized that the exact timing of the famines was probably not as predictable as the correlation might suggest, but that the important feature was the cyclical nature of the famine occurrence. However, in 1881, H. F. Blanford, the main Indian Government meteorologist, reported to the Famine Commissioners that, although he too believed solar heat varied cyclically, he had been unable to find a correlation of the sort that Lockyer had reported and there appeared no immediate possibility of using solar observations to predict weather in India. Later, Lockyer adopted another spectroscopic approach. He found from his observations of sunspot spectra that some lines certainly got wider at some part of the cycle, while some others got narrower and he found that the sun was hotter at the spot maximum and cooler at the minimum. Lockyer and his son, Jim Lockyer examined rainfall variations in the vicinity of the Indian Ocean as a function of Lockyer's spectroscopic estimates of the solar temperature. Lockyer explained the Indian data in terms of what he called 'heat pulses' (a positive pulse at maximum and a negative pulse at minimum). To both these solar heat pulses there corresponded rain pulses on Earth. Famines would tend to occur between these rain pulses, that is, between maximum and minimum in the spot cycle. This work was warmly welcomed in India[3].

All these factors contributed to pressurize the Government (or Secretary of State) by the Solar Physics Committee to establish a solar physics observatory in India.

## Mountain observatory

The dominant use of spectroscopy in the exploration of the physical nature of the celestial sources along with photography of the sky demanded bigger and bigger telescopes to gather more light and to push the studies to fainter limits and higher spectral resolutions. Such a need was also felt at the Madras Observatory in 1880s. In a letter dated 15 May 1882 (note 11) addressed to the Chief Secretary of the Government of Fort St. George, Norman Pogson states 'I have the honour to supplement one very important suggestion it obtained by the submission of a few further remarks upon desirability of establishing a branch observatory, equipped with a large modern equatorial at a hill station in South India. Twenty years ago, as stated in my letter referred to, an equatorial telescope of seven or eight inches aperture was regarded as ample for the scientific wants of a First-class observatory. The vast improvements in the manufacture of large object glasses have now, however rendered such instruments quite out of date, and nothing short of a twenty-inch glass can enable an Astronomer to keep pace with the enterprise of his continental and American rivals in science. The one observatory of India if deemed worthy of being maintained at all, should undoubtedly be equipped with a large telescope than it now possesses. At Melbourne, a forty eight-inch reflector has long been in use, at Sydney an eleven-inch refractor has been supplied, and at D'Urban in Natal, a large telescope is decided upon, while at Madras a very second class eight-inch is the largest available.

'A fact now generally recognized, though singularly over looked in the past times, is that by placing an instrument on a mountain, so as to eliminate several thousand feet of the grosser and lower atmosphere, its defining power is enormously improved, or in other words it is equivalent to a much larger telescope used near sea level.

'I would therefore, strongly urge that a large new telescope should be ordered with as little delay as possible, to be placed at a hill-station, as far above sea level as convenient; but whether the Nilgiris or the Palani Hills would be best suited for the purpose, I am not yet prepared to advise... .'

It is quite remarkable that Pogson could realize and emphasize the importance of locating the telescope at heights where better observing conditions prevail. He probably was influenced by the work of Piazzi Smyth on mountain observatories. Smyth was one of the two people who recommended Pogson for the Directorship of Madras Observatory (the other being John Herschel).

Although 'as early as 1717 Isaac Newton had recognized the limitation set on the performance of telescopes by atmospheric turbulence, and had suggested that the remedy for this "perpetual" was a most serene and quiet air, such as may perhaps be found on the tops of the highest mountains above the grosser clouds'[30].

'Isaac Newton's "serene air above the grosser clouds" lay the future for observations astronomy' is what was realized by one of the pioneers of observational astronomy and great champion of mountain astronomy Charles Piazzi Smyth, the 'peripatetic astronomer'. His experiments at the Peak of Tenerife in 1856 proved beyond any apprehension that the best image quality (seeing) transparency and spectral range are realized at the mountain tops compared to low altitudes where most of the observatories were located[29]. Charles Piazzi Smyth was one of the teachers to Michie Smith at Edinburgh (his B Sc degree certificate received in 1876 was signed by Piazzi Smyth, P. Tait among others). Michie Smith also continued to keep in touch with Piazzi Smyth even from India.

The first observatory that was consciously located on a mountain peak was the Lick Observatory on Mount Hamilton, where in 1888 the 36-inch Lick refractor was commissioned under the directorship of Edward Holden[30]. Indeed it was a bold and futuristic suggestion by Pogson to locate a branch observatory in Palani hills, barring the fact that no observatory in the British territories was located on mountain peaks till that time.

The idea of setting up a hill observatory in India for a limited number of years was first brought up in 1858, the year of publication of the Tenerife Report by Piazzi Smyth, but had to be shelved following the Indian Mutiny[29]. The plans were revived in 1861 by R. C. Carrington and W. S. Jacob, two astronomers who were on friendly terms with Piazzi Smyth. Carrington had in fact been on the Royal Society committee which advised on Piazzi Smyth's Tenerife expedition. Jacob was the Director at Madras Observatory from 1849 to 1858 and had his own private observatory at Poona before coming to Madras. At Madras, apart from the usual meridional astronomy Jacob specialized in measuring double stars and Saturn's rings, etc. Some of these results were reported by Piazzi Smyth at the Royal Society of Edinburgh. Jacob submitted a proposal for funding to set up a hill observatory near Poona at a height of 4000 feet to the Government after getting it endorsed by the Royal Astronomical Society. A grant of 1000 pounds was provided by the Treasury to put up a station for three years with a nine-inch Jacob's own telescope. Piazzi Smyth welcomed the project. Jacob after his return from England in 1862 to Bombay even informed Piazzi that he is proceeding to the mountain, but unfortunately died five days later[29].

When Pogson's proposal for a branch observatory to Madras (Observatory) to be located at a hill station in the South India was submitted, the Governor in Council of the Madras Government, agreed entirely and 'His excellency believes that no such large Equatorial as is proposed exists in the Peninsula and thinks it would be in the interests of science to equip such an observatory as is indicated. No officer could probably be found better qualified to start a new station than Mr Pogson' (15 March 1883 letter of communication from Chief Secretary, the Government of Fort St George (C. G. Master) to the Secretary to Government of India, Department of Finance and Commerce). The Astronomer Royal, W. H. M. Christie in a letter dated 14 April 1883 to the Government at Fort St George, re-specified the needs recognizing Pogson's proposal 'Taking then a general view of the requirements of astronomy at the present day, it would, I think, be desirable to maintain an astronomical observatory in India for the following definite objects. (a) Determination of accurate positions of the sun and a limited number of fundamental stars and so far as practicable, of the moon and planets. (b) Daily observations of the solar prominences and of other solar phenomena with the spectroscope.' Further elaborated by Pogson 'with regard to the subject of astronomical physics... if my suggestion for the establishment of a branch observatory on the Pulney Hills in the Madurai district, seven thousand feet above sea level, were carried out, such a position would, above all others, be the best for spectroscopic researches...' for both solar, stellar, nebular and other solar system objects.

The Governor, requested Michie Smith (Figures 4 and 7), 'a gentleman of considerable scientific attainments', who was a professor of physics at Madras

Christian College, also trained in practical astronomy and well acquainted with Pogson and the Madras Observatory, to undertake the site survey in Palani hills for a proper site for the observatory.

### Smith – early childhood

Charles Michie Smith was born on 13 July 1854 at Keig, Aberdeenshire, Scotland, seventh child of William Pirie Smith and Jane Robertson Smith. William Smith was a minister of Free Church till his retirement in 1881.

The early childhood of Michie Smith and life in a small town (village) in Scotland in Victorian Britain have been described by his younger sister Alice Thiele Smith[31] in a delightful book *Children of the Manse – Growing up in Victorian Aberdeenshire* (many of the details mentioned below are mainly from this source). Michie Smith was the third son in a large family of 11 children, four boys and seven girls (Figure 8).

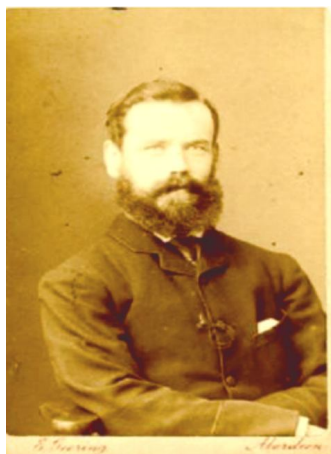

**Figure 7.** Michie Smith just before arriving in Madras in 1877.

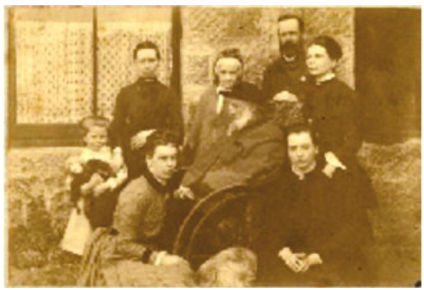

**Figure 8.** Family of Michie Smith during 1889–90 – Aberdeen. From left: Little Jeannie, Bella, Nellie, mother Jane, father W. P. Smith, Charlie, Lucy and Alice.

The elder two, William Robertson Smith and George, were thought to be highly talented students. George died young soon after he secured a place at Cambridge to study mathematics. William rose to celebrity at a fairly young age and became a professor and Thomas Adams Chair of Arabic at Cambridge in 1889. He was also the editor-in-chief of *Encyclopedia Britannica* for some time. As a student, he also worked as an assistant to P. G. Tait, professor of natural philosophy and a prominent physicist at the University of Edinburgh, helping him in teaching physics. Apparently, the famous author Robert Louis Stevenson was one of those who got benefitted by such Will Smith tutorials.

Charles Michie Smith (Charlie) was to follow his two brilliant elder brothers and inherited the legacy of being compared, unfortunately, with them, particularly by his father. The middle name Michie in Charlie's name comes from his paternal grandmother. Charlie was the most practical of all the children and had a special interest in both woodwork and astronomy. His sister writes that even as a boy 'Charlie, who was the family expert in woodworking, made fine hoops out of birch for the croquet game' and a cabinet to store them. He had a workshop in a secluded corner of the garden equipped with a large work bench and huge assortment of tools spending all his pocket money on it; of course, thoroughly disapproved by his father. His other interest was photography. 'Charlie, then, seventeen, had brought his fishing rod and could be safely trusted to go off alone to catch fish for tea. Best of all were the flounders that he simply caught with his hands while wading at the sea's edge.' He was taught at home by his father till he went to University; 'Curiously enough, Charlie too had suffered from his father's ire through being a poor classical scholar, though brilliant at Maths' which '…at times distinctly strained the relationship between father and son so that I think Charlie was pleased to leave home for King's College in Aberdeen' in 1870, which also coincided with his brother Will's appointment as Hebrew Chair at Free Church College in Aberdeen. Both brothers lived together in Willi's flat. After graduating, with M A at Aberdeen, Charlie went to Edinburgh to continue his scientific studies and ultimately getting his B Sc degree in engineering in 1876 (his degree was signed by Tait and Piazzi Smyth). He had the privilege of studying under such distinguished men as Archibald Geikie, P. G. Tait (father of J. G. Tait who taught in Central College, Bangalore) and Piazzi Smyth.

Soon after he found a job with a company manufacturing (laying) submarine cables across the world '…I can't be sure, but it's possible that Professor Tait and Sir William Thomson (Lord Kelvin), Willi's old friends putting Charlie's name forward to the company'. He was sent to the Caribbean to lay the cables, but returned to Edinburgh after few months because jobs became scarce and he was laid-off. Soon he got the appointment, from Free Church Foreign Missions Committee, as a professor of mathematics and physics at Madras Christian College. 'His departure was especially hard for all of us since it came at the start of Willi's long battle with Free Church. The two brothers corresponded regularly, of course, but Charlie could be of no direct help.' During these years and later, Charlie seemed to have extended lots of help, love and affection to his family, particularly to his sisters.

### In Madras – Madras Christian College

Michie Smith arrived in India in early 1877 at an age of 22 to take up the job in Madras Christian College (Figure 7). 'Life in India appeared to suit him eminently.'[31] He threw himself vigorously into his work both at College and University (Figure 9). He started the Christian College Magazine and contributed articles on a variety of science subjects ranging from development of practical electricity, meteors, volcanic eruptions, zodiacal light and even the developments at Madras Observatory. He took much interest in physical training of youth at the Presidency. He became the joint-secretary to the Madras Physical Training and Field Games Association which tried to encourage athletics in schools and colleges. He was a secretary to the Madras Literary Society for number of years. He was a member and President for several years of an organization called Friends-in-Need Society. He took up responsibilities as Fellow and later member of the Syndicate of Madras University for several years. He, obviously, was popular with students as evidenced by the letters students sent him when he ultimately left the college to accept the Directorship of Madras Observatory.

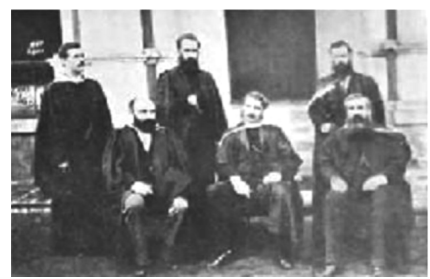

**Figure 9.** The Madras Christian College senate in 1880. G. Patterson is sitting on the left, William Miller (principal) in the centre and Charle Cooper on the right. Standing (from left to right) are A. Alexander, G. M. Rae and Michie Smith.

All this was done while he continued his own researches on meteorology and astronomy. He participated in the activities of Madras Observatory and was a great help to his friend Norman Pogson, the Government astronomer. He regularly communicated his researches to the *Proceedings of the Royal Society of Edinburgh* (note 12), which ranged from description of volcanic eruption and its consequences at Bandaisan in Japan, experiments regarding the surface tension of liquids, and spectra of vegetable colours using a direct vision spectroscope which was borrowed from University of Edinburgh. Of particular interest are his researches on zodiacal light. He obtained a grant from the Government (Royal Society) in 1882 for constructing a spectroscope specifically designed by him to observe and photograph the spectrum of zodiacal light (faint and extended). He built the spectroscope (with two interchangeable collimators and quartz lenses and dispersing prisms and slits) successfully and carried out visual observations at Madras. Attempts to photograph the spectrum were unsuccessful. '...very little colour can be distinguished in the spectrum, but under favourable conditions a distinct tinge of red can be observed. Except on few occasions the spectrum, as I have seen it, was continuous and quite free from bright lines, but on several nights in 1883 I saw what appeared to be a bright line.'[32] He continued his observations even at high altitudes of Doddabetta at 8642 feet above sea level in January 1885 '... in all my observations, which have been carried on at intervals since 1875, the spectrum has appeared continuous and free from bright lines except during the spring of 1883... the estimated position of the supposed line is wavelength 558, differs but little from the auroral line' (note 13). As he remarks '... zodiacal light on which more detailed and accurate observations are much needed is its exact position in the sky.' Michie Smith then quotes the work of Captain Jacob in Madras during 1856–58, who found that the axis of the light is inclined but slightly to the ecliptic. (It is presently known that zodiacal light is caused by interplanetary dust distributed in the zodiacal plane.)

Because of his researches in meteorology and astronomy, the Governor considered Michie Smith 'a gentleman of considerable scientific attainments' and invited him to conduct site survey in Palani hills for the branch observatory.

### Site surveys

Michie Smith conducted his site surveys for a hill observatory in two installments covering the period May to July 1883 at Palani hills and January 1885 at Doddabetta and its surroundings, as reported in his communications to the Governor of 7 August 1883 and 19 February 1885. His observations were mainly related to the general rainfall, cloud cover, interference due to mist formation, sky transparency, dryness (water vapour – by observing spectroscopically the water vapour lines near NaI D called the 'rain band'), haze (scattering due to dust in the atmosphere) and the stability of the atmosphere (seeing) both during the day as well as night. He conducted both spectroscopic observations of stars and Sun, zodiacal light, as well as visual observations measuring separations of double stars, their brightness, etc. and compared them with Madras (plains) and with each other of these sites. His conclusions were that the site in Kodaikanal, on the western side, at height of 7700 feet above sea level, with uninterrupted view of the horizons, is eminently suitable for a hill observatory.

A letter dated 9 December 1883, addressed to the Secretary, Government of India, Revenue and Agricultural Department, the Chief Secretary of Government of Madras, states '... invite attention to the accompanying paper, drawn up at the request of His Excellency the Governor, by Mr Michie Smith, a gentleman of considerable scientific attainments, who has made a series of careful observations on the Palani hills in the Madura district. His excellency the Governor in Council would again urge very strongly the advisability of establishing a branch observatory in these hills, where a large portion of clear nights may be confidently reckoned on during the year, and where, owing to the absence of dust and haze, the conditions are most favourable. There can be little doubt that a succession of observations with a powerful equatorial placed on these hills would lead to very valuable scientific results.'

The Government sought advice of the Astronomer Royal, W. H. M Christie, on the location of a large (about 20-inch) equatorial in a branch observatory in Palani hills. His advice (letter of 17 August 1883 to Under Secretary of State for India), surprisingly was not either based on the merits of the site or the astronomical needs or feasibility of installation, but because Pogson has lot of accumulated unreduced observations which need to be reduced and published; '...it does not seem to me desirable to charge him with additional work of supervising a large equatorial. On Mr Pogson's retirement the question of establishing a branch observatory or removing Madras Observatory to a more favourable station might be considered. ... As regards the advantages of a mountain station for astronomical observations the evidence is somewhat contradictory, and I am not prepared to endorse Mr Pogson's remarks as to the enormous improvement in defining power necessarily resulting from a great elevation. It might well happen that better results would be obtained at a moderate altitude or even at the sea level, if the climate were fine and healthy. I am, however, not sufficiently informed as to the climate of Madras to be able to offer my opinion on the expediency of moving the observatory to a neighbouring station.' As a result, the decision to put up a large equatorial was shelved (or indefinitely postponed) till Pogson's retirement (note 14) even before receiving Michie Smith's second report.

Meanwhile, the Indian Observatories Committee was constituted and it started harassing Pogson about publication of earlier accumulated meridional observations. Pogson became ill in June 1891 and was diagnosed as suffering from virulent cancer in the liver of rapid growth. His commitment to astronomy was such that he immediately tried to put his affairs in order and even wrote a letter to the Government about his possible successor. The letter of Pogson dated 20 June 1891 to Chief Secretary of the Government (just before his death) states 'The individual who is most intimately acquainted with the details and (steps) of my (past) labours in the Madras Observatory is Professor C. Michie Smith, B Sc. He is a Fellow of the Royal Astronomical Society and of the Royal Society of Edinburgh and has for several years past been entrusted by the Royal Society of London with a large and valuable spectroscope for researches on the Zodiacal light. I may add that he has undergone a course of training in Practical Astronomy under Dr Copeland, Astronomer Royal of Scotland. I would therefore urgently advice and solicit as my dying wish that the said Professor C. Michie Smith might be appointed my successor as I am convinced that if this were alone (will) my unfinished works would be completed as closely as possible as if I had lived to finish them myself... .'

*Michie Smith – successor to Pogson at Madras Observatory*

Pogson died on 23 June 1891. A letter (Public Department No. 14) addressed to the Secretary of State for India, the Madras Government states 'We have appointed to the vacancy thus created, Mr C. Michie Smith, B.Sc., F.R.A.S., F.R.S. EDIN., who is a well known observer and is employed as a professor on the staff of the Madras Christian College. This gentleman has consented in addition to carrying on the current work of the Observatory, to proceed with and conclude as early as possible the publications of the late Mr Pogson's observations. We may mention that this arrangement has been adopted in accordance with a special wish expressed by the late Mr Pogson … There being no native of India competent to perform the duties of Government Astronomer in satisfactory manner and it being imperatively necessary to take immediate steps for carrying on the work of the Observatory...'. He was appointed as officiating Government Astronomer.

Michie Smith's immediate job soon after moving to Madras Observatory was to attend to the pending reductions of the meridional observations and bring them out as publications '*Results of Observations of the Fixed stars Made with Meridian Circle at the Government Observatory in the year 1871 to 1882*'[33] and later the '*New Madras General Catalogue of 5303 stars for the epoch of 1875.0*'. He successfully managed to bring out all the pending observations in four volumes, last two were published under the joint names of Pogson and Michie Smith.

### Establishment of the Kodaikanal Observatory – plans, buildings

Pogson's death in a way reactivated the proposals for an astronomical physics observatory. John Eliot, the Meteorological Reporter to Government of India was to take stock of the situation in Madras Observatory. He was well aware of the recommendations and requirements of the Solar Physics Committee in India as well as the necessity of reorganizing the Madras Observatory by Madras Government in the context of a mountain observatory and the comments of Astronomer Royal in this regard. He tried to synthesize a solution to various requirements and wrote a letter on 13 October 1891 to the Government of India (which was circulated to the Governor of Madras for comments). He felt 'The present position at the Madras Observatory affords a favourable opportunity of considering the future status of one large Astronomical Observatory in India.' According to his information, the secretary of State has decided, on the advice of Astronomer Royal, that it was necessary to maintain an Astronomical Observatory in India that would conduct astronomical physics, including daily monitoring of solar features on and off the disk, by photography and spectroscopy in addition to the astronomical work that has been routinely being practised at Madras Observatory of determination of fundamental positions of stars. The Government of India accepted this outline of work as a basic requirement, in addition to other recommendations. The Madras Observatory's aspirations included, in addition to positional astronomical work that was going on, as 'Astronomer Royal has pointed out the necessity of the addition of observations in connection with astronomical physics. None have been hitherto taken at Madras, although Mr Pogson recognized their importance, first, because the atmospheric conditions at Madras were not suitable, and, secondly because Mr Pogson was so pressed with work and arrears of work that he had no time to introduce these observations.' As to the requirements of Solar Physics Committee, the actinometric work that was started first at Leh was later shifted to Mussorie and then to Simla. The daily photography of Sun is being done at Dehra Dun thus far had to be continued elsewhere. The need to conduct solar spectroscopy at a hill station observatory was felt necessary, as it had not even started. 'Messers Blanford, Pogson, Hennessey and others have urged the necessity of this for solar physics observations. Mr Pogson has also stated that much of the astronomical work could be done at a hill station than at Madras.'

About the location of the hill observatory, North Indian sites of Leh, Mussorie and Simla were considered and rejected because of the dust in the atmosphere in these locations[34] (note 15). The South Indian hill stations were seen to have much less dust and cloudiness than their counterparts in western Himalayas. The choice then came down to two places, Kodaikanal in Palni hills and Kotagiri in the Nilgiris. The Government of Madras was instructed to request the officiating Government Astronomer at Madras, Michie Smith to undertake further detailed investigation of these two places for a period of a year.

Another suggestion Eliot made was regarding the organization of the observatory. The proposal was to establish an Astronomical and Solar Physics Observatory at a hill station in southern India with a staff of two European scientists and native assistants and to reduce the Madras Observatory to a branch observatory of the hill observatory, with the assignment of carrying out meteorological, magnetic and time signal-related observations mainly with native assistants and a half-time officer to supervise. The control would be with Government of India with advice from the Observatories and Solar Physics Committees. The astronomers were to be selected in England by the Observatories Committee. He even recommended Michie Smith to head the Observatory and supervise the Solar Physics section as well.

By 5 August 1892 Michie Smith could complete his investigations of the two prospective sites for the hill observatory and submitted a report for the third time regarding their relative merits and compared them with observations at Madras. The observations conducted were: (a) of the Sun with special reference to spots, and faculae and their spectra, (b) Venus and Saturn, (c) star clusters and nebulae, (d) of double stars, (e) stellar spectra, (f) measurements of stellar magnitudes and (g) photographs of star trails. The period covered was between February and April. After evaluating the meteorological and astronomical observations, 'A discussion of the results by Mr J. Eliot, Meteorological Reporter to the Government of India, and myself showed clearly that Kodaikanal was to be preferred. ... the result left no doubt that Kodaikanal was far superior in every respect.'[35]

Eliot's recommendation along with Michie Smith's site survey report were considered by both the Solar Physics Committee and the Indian Observatories committee. The Solar Physics Committee, with Norman Lockyer participating, endorsed on 22 July 1892 all the suggestions of Eliot, and the hill site, but noted in case of difficulty in establishing the astronomical observatory in its entirety, the first priority should be to establish the Solar Physics Observatory with related activities with emphasis on continuing, without a break, solar photography that was being done at Dehra Dun. They also commented that 'experience already gained at South Kensington in conduct of the work in solar physics shows conclusively that no large initial expenditure on buildings is essential'. About appointment of Michie Smith as the Director of Solar Physics Observatory, the committee expressed that 'we are not aware of there being any one better qualified for the post at present in India'.

The Indian Observatories Committee, chaired by Lord Kelvin, resolved on 26 October 1892 that the 'Committee highly approves of the proposal to establish an Imperial Observatory for India on a site to be chosen in the highlands of Southern India' and also that 'the Committee, from personal knowledge of Mr. Smith's ability (note 16) and attainments are of opinion that he would be very suitable Superintendent of the proposed Solar Physics Section of the Imperial Indian Observatory'.

In another communication dated 4 March 1893, Eliot after visiting Madras and consulting the Government as well as people of Madras Observatory, and in accordance with recommendation of Famine Commission, made a detailed and elaborate report to the effect that establishment of Solar Physics observatory at Kodaikanal is to be taken up immediately for 'greater benefit to science', whereas establishment of the astronomical branch could be delayed by five years, till that time they (i.e. astronomical observations) would be carried out at Madras. He also conveyed that the Madras Government would welcome handing over the observatory to Government of India and expressed strong opinion about Michie Smith that 'he was by his special scientific attainments and knowledge of the conditions of scientific work of India, peculiarly fitted for the appointment of Director of the Solar Physics Observatory, and that his services in carrying out the publication of the Madras Observations had been especially valuable and deserved full recognition'.

Finally, Secretary of State for India in a letter dated 31 August 1893 to the Governor-General of India, states that after receiving an endorsements from Indian Observatories Committee through the Astronomer Royal's letter of 31st July and the Solar Physics Committee through the letter dated 2 August 1893 from its secretary, Colonel Donnaly, on the proposal of establishing on a small scale a new observatory at Kodaikanal in the Palani hills (under the Direction of Michie Smith) 'I sanction your proposals, which involve an initial outlay of Rs 25,000 (note 17) and yearly increase of Rs 1689', and he further invited the attention of the Government to the suggestion of 'continued photographic registration of sun spots': and 'Michie Smith's consulting the Astronomer Royal regularly' regarding the catalogues of the old observations. Thus a formal approval was given and an order dated 21 November 1893 communicated to Michie Smith for arranging the budget, estimates and plans required by Government of India. The control of Madras Observatory and the funds connected with it were transferred to the Government of India from 1 April 1894, placing the organization and management of the new Solar Physics Observatory under the Meteorological Reporter of the Government of India.

Soon after his appointment, Michie Smith started planning for establishing the observatory; its layout, domes, etc. (his letter of 27 November 1893 to Astronomer Royal). The initial building plans of the observatory and the instruments being acquired have also been described along with detailed account of site selection by Michie Smith in an article written on 19 February 1895 in *PASP* at the invitation[35] of Holden of Lick Observatory. 'The proposed equipment consists chiefly of instruments already in India. These are two photoheliographs of KEW pattern, with a 4-inch and 6½-inch objectives, giving solar pictures of 8-inches and 12-inches diameter, at present at Dehra Dun; a 6-inch Cooke equatorial, with a powerful, but somewhat obsolete, form of prism spectroscope for solar work, and the necessary minor instruments. In addition to these, there will probably be a 6-inch photographic lens of 36 inches focal length, mounted with the LEREBOURS and SECRETAN equatorial of the Madras Observatory as a finder, a small spectroscope, with a Rowland grating fitted for photography, and the zodiacal light spectroscope made for me by HILGER some years ago.' Further, regarding the buildings 'The final designs for the buildings have not yet been prepared, but the observatory will probably be built in the form of a cross, of which the longer arms will lie east and west, with an equatorial dome at the east, and a transit-room at the west extremity. The north and south arms will also terminate in domes for one of the photoheliographs and the photographic equatorial. For spectroscopic work, probably, a heliostat will be used in connection with a fixed telescope, the axis of which is pointed to the pole. The Astronomer's house is to be built a little below the observatory, so as to be sheltered from the strong winds on the hilltop... .'

The first structure to start at Kodaikanal Observatory, after acquiring the land (89 acres) was the astronomer's house, for which foundations were laid at the end of April 1895 (CMS letter to Christie of 25 April 1895).

Michie Smith was granted leave on duty to visit England for few months in mid 1895 and consult both the Committees about the developing plans regarding instruments, science and buildings of Kodaikanal Observatory. At the invitation of Lord Kelvin, the Chairman of Indian Observatories Committee (note 18), Michie Smith presented his plans for Kodaikanal before the committee on 17 July 1895. The meeting was initiated by the Secretary of State for India, Lord George Hamilton. The committee approved 'generally of the plans laid before it for the Kodaikanal Observatory, and considers that they are well adopted for the work to be done. In particular, it approves of the proposals regarding the solar spectroscope, and strongly recommends that an instrument of this form be provided'. Subsequently, the proposals were sanctioned by the Secretary of State (letter of 12 August 1895 to Governor at Fort St George, Madras Government order dated 2 September 1895). The new solar spectroscope proposed consists of lens of 40′ focus and 6-inch aperture, slit, concave grating and apparatus for both viewing and photographing the spectrum.

In October 1895 the foundation stone of the new observatory was laid by the Governor. By July 1897, the north–south line for the observatory building was marked by Michie Smith and construction plans were handed over to the Government architect. Soon it was time to plan for the 1898 total solar eclipse that was to pass over central and North India.

### Total solar eclipse of 1898

After the historic 1868 total solar eclipse, the total eclipse of 22 January 1898 occupies an important place in the context of solar physics in India. This is the eclipse in which detailed and elaborate instrumental set-ups have been deployed by several international teams that include one headed by Naegamvala of Takhtasingjii Observatory, Science College, Poona, and another by Madras Observatory led by Michie Smith, at several places along the totality path that passed through the states of Maharashtra and Madhya Pradesh. John Evershed, who was to join Kodaikanal Observatory, had also come as part of the British Astronomical Association delegation. Delegations from institutions belonging to different countries including England, America and Japan participated in observing this eclipse. Observational camps were established all across the country from Vizayadurga on the west coast, where Norman Lockyer was stationed, to Jeur where Naegamvala, William Campbell of Lick Observatory and others set up their camps, to Talni where John Evershed had his equipment, Shadol, where Michie Smith, The Astronomer Royal, H. H. Turner and others had their camps, Dumraon where St Xavier's College, Calcutta had its camp and several other places between and beyond.

'The observations of the total solar eclipse in India has been a magnificent success' declared Maunder[36] on the day of the eclipse. All the stations on the totality path had clear skies. The spectacle was further described by Maunder, stationed at Talni 'we looked up at the magnificent spectacle before us. The darkness did not equal that of the eclipses of 1886 or 1896, but the corona stood out in the sky as a vast silver star, bright and more extended than when I saw it 11 years before. Two fine leaf shaped extensions stretched out almost horizontally east and west, whilst nearly, but not quite, on the sun's equator, directed south-west, was the greatest ray of all, two millions of miles in length almost, pointing to where one celestial brilliant glittered several degrees away' (note 19).

The main concerns of the Kodaikanal Observatory team were 'changes in the conditions of the Sun's atmosphere associated with changes in the number of sunspots'. This eclipse is to take place near the time of minimum sunspots and would provide good comparison to the eclipse results obtained at maximum sunspot numbers. The emphasis was on photographing the Corona at different scales, with a 40 foot focus and 6-inch aperture lens as well as with a 5 foot focus and 4-inch aperture lens. The 40-foot telescope gave images with 4 and 3/4 inch diameter moon on 18″ × 18″ plates, whereas the 5-foot photoheliograph gave an image of 0.6-inch diameter moon and could trace the corona to about two solar diameters or more (Figure 10). The results showed the extreme brightness of the inner corona and the rapidity with which it fades away with increased distance from the limb; the evident connection between prominences and certain details in the corona; the striking resemblance between the corona of 1896 and 1898 (ref. 37). The short, half second exposure given with the great 40-foot telescope 'shows the very beginning of the eclipse with a range of small prominences round nearly half the sun's limb'. The striking long streamer is conspicuous. The team included R. L. Jones, Surveyor-General, J. L. VanGeyzel, M. R. Ry. K. V. Siva Rama Aiyar, Assistant to the Government Astronomer (Michie Smith) and others.

The emphasis of the experiments of both Evershed and Naegamvala was on spectroscopic observations of the corona, chromosphere, prominence and more importantly, the nature of the 'flash' spectrum (Figure 11).

The dark Fraünhofer lines in the solar spectrum are thought to have been formed in the reversing layer by cooler gases that absorb the sun's continuous spectrum. In the eclipse of 22 December 1870, C. A. Young watching the spectrum of 'the dwindling arc of sunlight with a slit spectroscope, as the moon had all but hidden the sun, saw at the moment of second contact the ordinary solar spectrum with its dark lines on the continuous background disappear and then all at once, as suddenly as a bursting rocket shoots out its stars, the whole field of view was filled with bright lines, more numerous than one could count. This beautiful appearance has since become known as the "flash" '[38].

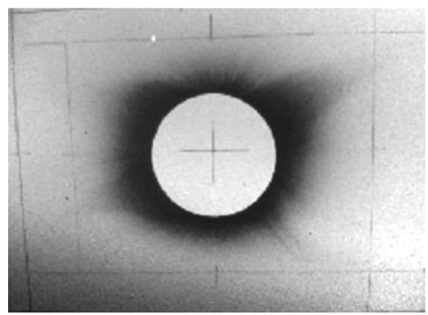

**Figure 10.** Photograph of corona at total solar eclipse of 22 January 1898 as observed by Michie Smith with 5-foot photoheliograph at Shadol.

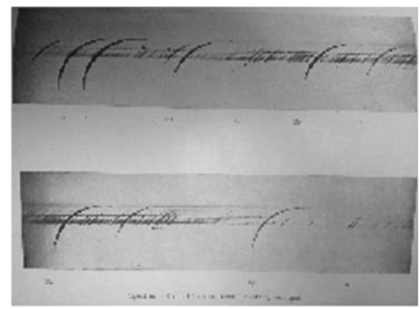

**Figure 11.** Flash spectrum in the 1898 total solar eclipse as observed by Naegamvala from Jeur. Bright regions are shown as dark and vice versa.

Evershed intended to photograph the spectrum of the reversing layer and the emergence of emission spectrum from the Fraünhofer dark line spectrum. He managed to obtain spectra extending into the ultraviolet up to 3346 Å. 'A new feature was shown in the hydrogen spectrum of prominences, consisting of a continuous spectrum beginning at the limit of the hydrogen series (Balmer series) of lines at $\lambda$ 3646 and extending to the end of the plate. This proved to be the counterpart of the continuous absorption spectrum discovered by Sir William Huggins in the stars having very strong hydrogen lines.'[2] The measures of wavelengths and intensities of lines of 'flash' spectrum showed that 'flash spectrum represents the higher, more diffused portion of the gases which by their absorption give the Fraünhofer dark line spectrum, and the flash spectrum is the same in all solar latitudes. Previously, it was considered that the flash and the Fraünhofer spectra were separate and distinct phenomena'[2,39]. Evershed also reconfirmed the wavelength of coronal line 1474K as 5303 Å.

Jeur, in Satara district of Maharashtra, attracted several astronomers as the camp site for observing the 1898 total eclipse of the Sun. Naegamvala's team from Maharaja Thakhtasingji Observatory and Science College, Poona had an elaborate set-up, with 'greatest number of assistants upon the ground, was helped by the Principal, some of the professors of the college, and other gentlemen, and a strong contingent of native students whom he succeeded in knocking into shape by constant drills for a week in advance'[38,40]. W. W. Campbell of Lick Observatory, California had a neighbouring camp, with a large (longest) telescope of 40-foot focus, the end of which was located in a pit 8-foot deep, dominating the site. His wife, Elizabeth Campbell[41] (note 20), organized the running of the camp for him. Japanese astronomer Terao and his team were also at the same site.

The major Indian effort at observing the eclipse was that of Naegamvala[42]: 'From the first I had proposed to concentrate the powers of the expedition in the spectroscopic line, a secondary place only to the subjects of photographing the corona and of the eye-observations.' Particularly, the phenomenon of 'flash spectrum', in the words of Maunder: 'It was no exaggerated sense of its importance which, in recent eclipses, has placed photographs of the "flash" as absolutely the most instructive observations that could be made.' Naegamvala's largest and most important instrument was a large objective prism mounted upon a very heavy substantial equatorial stand, upon which it moved by clockwork, for obtaining the spectra of 'flash' and corona. Another instrument for spectrum photography worked with a 12-inch siderostat, while an integrating spectroscope, working direct, was manipulated by hand. Naegamvala presented the results of his several experiments in an impressive detailed report published in 1902. The delay in publishing the report is due to as he explains, 'Though Mr W. Shackleton had secured the first record of "flash" at Novaya Zemyla in 1896, the spectrum of this important solar phenomena, was for the first time adequately obtained at the Indian Eclipse of 1898, and it was at once evident that proper appreciation of this important record could only be gained by comparing it with stellar spectra of solar and other related types', which he tried to obtain for the previous two years without much success. His report deals in detail about the spectra of the flash (Figure 11), which he compared with the Fraünhofer spectrum and came to the conclusion similar to that mentioned by Evershed. To the question 'Is there a definite layer of gases in which, if not all, the vast majority of the Fraünhofer lines are represented?', the answer is affirmative. The prominence spectra and the wavelengths of the three coronal lines present in his spectra (Figure 12), etc. were discussed in detail with respect to each instrument used. He also produced plots showing the variation of the shape of corona with respect to sunspot numbers over the years and showed how it changes with the solar cycle. This work of Naegamvala might be considered as the highlight of all his astronomical achievements.

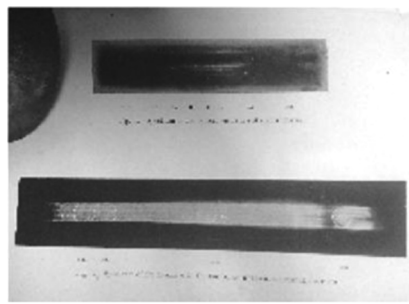

**Figure 12.** Slitless spectrum of corona in the 1898 eclipse as observed by Naegamvala. Bright rings are the coronal lines.

### Michie Smith–Lockyer feud

Michie Smith laid the north–south line for the building on July 1897 and handed over the work to the consulting Government architect in December hoping that 'if no unforeseen trouble arises, the buildings would be ready to receive instruments in less than six months' (Annual Report of the Observatory, 1897–98). The unforeseen trouble did arise in the form of Norman Lockyer.

In the Annual Report of 1898–99, Michie Smith writes 'the Government of India requested the Astronomer Royal, and Sir Norman Lockyer to report on the various Indian observatories. The former after visiting Kodaikanal approved generally of the plans for the observatory there and made some suggestions for minor alterations (dome diameter to be changed from 15 feet to 18 feet, etc.) which were at once adapted (note 21). Sir Norman Lockyer, on the other hand, without visiting the place (spent some hours in the drawing room of the Astronomer's house in Madras) objected entirely to the plans and on his return to England represented to the Secretary of State for India that the buildings were "too costly and too permanent" and were badly designed and unsuited for their purpose. He went on to point out that "the South Kensington Solar Physics Observatory thus equipped with temporary structures is the most powerful in the world. It does more and better work than the similar institution at Potsdam where the buildings cost £250,000" and urged that the new buildings at Kodaikanal should be like those at South Kensington "shanties", built of wood and canvas. As a consequence, the Secretary of State telegraphed out that the work on the observatory was to be stopped till the reports of the Astronomer Royal, and Sir Norman Lockyer had been fully considered. To anyone acquainted with the climatic conditions existing in Kodaikanal the proposal to house valuable instruments in such "shanties" as Sir Norman Lockyer recommends seems as strange as his estimate of the relative values of the work done at South Kensington and Potsdam and the Government Astronomer protested against this proposal (since this was written the roofs of two of the houses, though well fixed, have been blown off. The corrugated iron being in parts torn off as if it had been little stronger than card board). Whether or not this protest was forwarded to the Indian Observatories Committee is not known, but the result of the deliberation of the committee was that no reference what ever was made to the buildings after a delay extending from the beginning of June to the end of October according to the designs which have been so strongly condemned'. The result was a long delay in completion.

Some antagonism was built up between Michie Smith and Lockyer. In a letter dated 12 May 1898 addressed to Christie, the Astronomer Royal, Michie Smith complains '... Lockyer wished to turn the whole of the Kodaikanal buildings with a series of "shanties" and now you probably know that he has got the Secretary of State to telegraph out to stop all the buildings except the two domes. …Of course I know that Lockyer's action is taken mainly out of spite against me & I am sure that if the Secretary of State had known of the position he has taken up ever since my appointment he would have asked your advice before taking this action... .' The 'spite' which Michie Smith mentions in his letter might partly be due to the fact that Michie Smith belonged to the school of Piazzi Smyth and lacked proper training in solar physics.

The reasons for Lockyer's attitude might be because he was so thoroughly impressed by the work of his 'former student'[3], Naegamvala, 'who, so far as I know, is the only person in India particularly familiar with Solar Physics work' that he was unable to appreciate the effort put up by Michie Smith at Kodaikanal. He visited Naegamvala in Poona before he arrived in Madras. 'Mr Naegamvala was good enough to show me the results he had obtained (of the Total Solar Eclipse) and I found them of a high order of excellence, while the programme of work he undertook showed a large grasp of the various solar problems which await solution at such a time' (quotes from Lockyer[16]). While he saw an enthusiastic effort being put up at Poona, where he was not expecting any, on the contrary, at Kodaikanal where Lockyer was expecting to see lot of progress, he did not find enough to his liking. As a result, Lockyer had apparently lot of criticism about the functioning of Kodaikanal. As was reported in *Indian Engineering* '... Much fuss was made about the transfer of the 6″ Cooke Refractor to Madras for carrying out the new programme; we learn that this happened in 1894, and, inexplicable as it seems to us, we note from the report of Sir N. Lockyer, that the instrument on arrival was put in the godown and the beautiful spectroscope with which he had made some of his first observations of "widened lines" in sun-spots *"was destroyed at an expense of £65 under the guise of reconstruction"*. This is indeed a sad tale. But what is worse; a new instrument purchased ostensibly for conducting the observations required by the Solar Physics Committee at home is declared by the same high authority to be "*quite useless*" for the purpose ... The letter of remonstrance that Sir Norman Lockyer wrote to Mr J. Eliot before leaving India is in itself a sufficient commentary on the manner in which matters have been allowed to slide at Madras'.

While Lockyer was keen on solar physics work to start at Kodaikanal as quickly as possible with as little investment as possible (the idea he was harping on with Solar Physics Committee since 1893), Michie Smith, on the other hand, was concerned more with establishing a proper observatory. In an earlier letter to Christie, Michie Smith remarked about Lockyer's intentions 'my interview with Lockyer was trying but we both kept our tempers. His chief object was to try and get me to at once start spectrophotographs of the Sun on the Hale-Deslander plan with an apparatus made up of odds & ends from other instruments set up in a "shanty" "which would not cost £5". I pointed out that I had neither the necessary apparatus nor at present the time to take it up on which he offered to lend me his son or one of his assistants from South Kensington! I did not jump at the offer. He would like to turn Kodaikanal Observatory into a series of "shanties" (which Michie Smith abhorred).' Moreover, 'About this time (about 1900), the chance arose for Lockyer to go to India to take charge of the new solar physics observatory at Kodaikanal, but he finally decided to stay at South Kensington.'[3]

Anyway, as it turned out, the feud seemed to have died quietly and things continued (maybe through the intervention of Astronomer Royal) at Kodaikanal with some delay. By March of 1899, 1000 coolie loads of books and instruments were transported to Kodaikanal with remarkably little damage and Michie Smith took up his residence by the end of February (Figure 13) to supervise (advise) the construction of buildings and to receive instruments on arrival (Annual Report of the Observatory 1898–99). Officially, the observatory started functioning from 1 April 1899.

### Astronomy (solar physics) at Kodaikanal

'During the first few years, the work of the Director was naturally concerned mostly with lay-out, planning the organization. Prof. C. Michie Smith the first Director of the observatory did much pioneering work in this direction. The constructional works were done under his personal supervision. Planted trees for improvement of the solar image, installed the instruments for routine observational work and formulated daily observations. During a short period when Michie Smith was absent on leave, Observatory was under the direction of Mr C. P. Butler'[43].

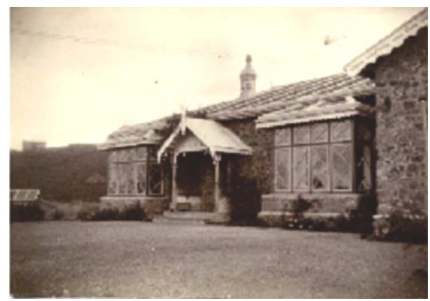

**Figure 13.** Astronomer's residence in Kodaikanal Observatory premises, 1900(?).

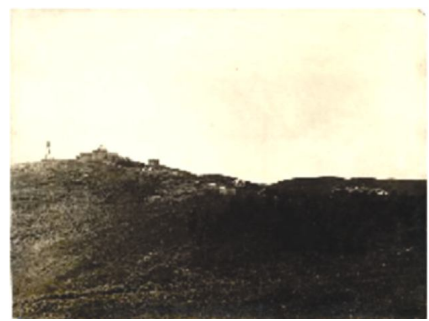

**Figure 14.** Kodaikanal Observatory probably in 1905, certainly earlier than 1908.

Systematic solar observations commenced at Kodaikanal from early 1901. Magnetic, meteorological and seismological observations were in progress at Kodaikanal from the very beginning.

Michie Smith seemed to have made a visit to England in 1900 June–August to consult the Indian Observations Committee (and Astronomer Royal) and also visited Aberdeen to see his family. Soon after his arrival back in India, he sent details of his plans for a Hale-type spectroheliograph, about which he had consultations with Hale and new spectroscope for sunspot and prominence work to the Astronomer Royal on 24 September 1900 and sought his advice. In 1902, a spectroheliograph to image in Ca ii-K line was ordered from Cambridge Scientific Instruments Company. It got completed and arrived at Kodaikanal in 1904. The instrument consisted of a 12-inch achromatic triplet with a focal length of 20-feet and was fed by a 18-inch silver-on-glass mirror Focault siderostat. The image of the Sun and the camera remain stationary, while the total instrument consisting of collimating lens, camera lens, prisms and slits fixed on a rigid frame moves on rails perpendicular to the optical axis (across the Sun's image).

In 1903, a spectrograph was built using a diffraction grating and a couple of existing lenses and used with a 6-inch Cooke equatorial to obtain spectra of solar features. 'A three-prism Evershed spectroscope was also ordered from Hilger Co. (the same company which provided Michie Smith his zodiacal spectroscope), for the use with the same setup which arrived in 1904. This gave excellent spectra of sunspots and prominences. One of the significant piece of work done at this time was observations of 'D3 (He i) as a dark line in the solar spectrum'[44], based on spectra obtained around the sunspots during one year period of March 1904 to March 1905. Most of these observations were made by M. Sitarama Iyer and G. Nagaraja Aiyar, the second and third assistants. By 1896, Michie Smith employed fairly well-educated persons as assistants like K. V. Sivarama Aiyar (with an M.A. degree), Sitarama Iyer (with a B.A. degree). They arrived at the conclusion that 'as a general rule, it may be said that the more disturbed the spot greater is the probability of D3 appearing as a dark line, although several occasions exist where it was seen in the spots that were quite as shown by the C (H$\alpha$) line'.

The 6-inch Lerebours and Secretan equatorial is modified and used as photoheliographs to obtain daily white light pictures of the Sun. (This is one of the oldest telescopes in the country, and still in use[19].) Regular series of scientific publications from the observatory was started by Michie Smith in 1904 as *Kodaikanal Observatory Bulletins* (*KOB*). The first paper in this series was by Michie Smith on the widened lines in sunspot spectra.

The daily solar observations made were both visual and photographic of prominence spectroscopy, with photoheliographs and spectroheliographs in the K line (and H$\alpha$ after Evershed added H$\alpha$ spectroheliograph in 1914). Positions of sunspots and prominences were recorded and marked on charts.

Michie Smith visited George Hale at Mt Wilson, with whom he was in constant touch, in early 1904 to discuss in detail the developments in solar physics and the kind of work Kodaikanal Observatory should be engaged in coming years.

*Arrival of Evershed*

Nominations for the position of an assistant to Michie Smith and also a possible successor to him as Director of Kodaikanal, after his retirement was solicited by John Eliot, the then Director General of Indian Observatories, from Astronomer Royal and the Indian Observatories Committee in 1902. The position was to be filled-in in due course. John Evershed, by then a well-known practitioner of solar physics and an 'irresponsible amateur' (as he describes himself) also expressed interest in 1902 along with few other candidates. 'In the year 1905 it was largely through his (Sir William Huggins) influence that the India Office offered me the post of assistant to Mr Michie Smith, director of the Kodaikanal Observatory.'[2]

John Evershed and the solar physics at Kodaikanal have been described in great detail in the IIA Archives Publication[45] and by Hasan *et al.*[46], commemorating the centenary of the discovery of the 'Evershed effect'. As such, we present in the following section, mostly his (and the observatory's) non-solar astrophysics and some highlights during his stay at Kodaikanal from 1906 to 1923 (which covers the period till Michie Smith's death).

After getting married in 1906 September, Evershed and his bride left England for Kodaikanal by way of America, visiting observatories and meeting astronomers there, finally reached Kodaikanal on 21 January 1907. 'My early work at Kodaikanal was largely concerned with Cambridge spectroheliograph which I brought into working order in early 1907; and so began the long series of photographs of the Sun's disk and the prominences.'[2]

Few months after his arrival, Comet (Daniel) 1907d showed up in the sky during the southwest monsoon period at Kodaikanal. In spite of the very poor weather, Evershed decide to obtain the cometary spectrum and devised the instrumentation efficient enough to get spectra in a short time (the comet could be seen for at most three-quarters of an hour). He devised a prismatic (objective prisms) camera with two 60° prisms of crown glass and photographically corrected lens attached to 6-inch Cooke equatorial in the south dome. He could get three occasions and obtained spectra of the comet on two with exposures of half-an-hour each. He used the spectra of bright stars Procyon and Sirius obtained on the same plates to get the wavelength calibration. The spectrum covered the region from H$\beta$ to ultraviolet shortward of $\lambda$ 3570. This was the first ever ultraviolet spectrum of any comet observed till that time. He could see bright monochromatic images of the tail (tail bands) at wavelengths $\lambda\lambda$ 3580, 3690, 3780, 4010, 4260, 4550, which were not very bright in the nucleus of the comet. These bands extended to 0.5–1.5° from the nucleus. Identification of these was not known at that time[47]. These are now known as the famous 'comet-tail bands' due to $CO^+$ ($A^2\Pi - X^2\Sigma^+$ system) and later seen in many other comets[48,49]. Evershed is the discoverer of these tail bands of $CO^+$ in UV in comets (although not acknowledged).

According to Evershed[2], 'One which was destined to occupy my thoughts for the rest of my life was a spot on the Sun's disk'. The year 1909 was an eventful year in the history of Kodaikanal Observatory, because of the discovery of 'Evershed effect', followed by finding one of the spots with highest strength of magnetic field of 10,000 gauss. These discoveries came about because 'the observatory possessed an excellent grating (a 6-inch plane grating by Michelson) of about 70,000 lines, and with this I constructed a high dispersion spectrograph'.

He used the high-dispersion spectrograph to find out whether there was any motion of ascent or descent in sunspots. 'An excellent opportunity presented itself for this work on 5 January 1909, with two large spots on the Sun, and excellent definition after heavy rainstorms. The spectra revealed a curious twist in the lines crossing the spots which I at once thought must indicate a rotation of the gases, as required by Hale's recent discovery of strong magnetic fields in spots, but it soon turned out from spectra taken with the slit placed at different angles across a spot that the displacement of the lines, if attributed to motion, could only be due to a radial accelerating motion outward from the centre of the umbra (Evershed effect). Later photographs of the calcium lines H and K and the hydrogen line $\alpha$ revealed a contrary or inward motion at higher levels represented by these lines.'[2]

A great sunspot in September 1909 showed a spectrum in which the Zeeman splitting of the iron lines in the region of H and K gave a magnetic field strength of 10,000 gauss during a 'flare'[50]. 'The accompanying magnetic storm upset the Indian Telegraph system, for which we appeared to be held responsible by the Director of Telegraphs'[2] (note 22).

Both Michie Smith and Evershed observed the Halley's comet from mid-April to mid-May of 1910, which put up a memorable display with its tail extending for 100° far up towards zenith. Both spectroscopic and photographic observations were systematically done to see the changes in its appearance[51] over the period (Figure 15). These revealed an accelerating outward movement of the tail.

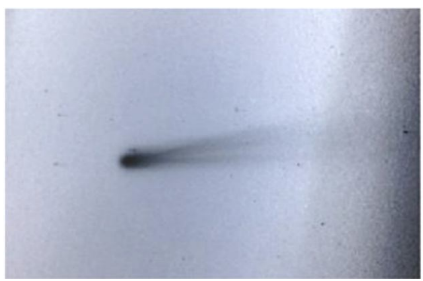

**Figure 15.** Halley's comet observed at Kodaikanal on 3 May 1910.

'It appeared as though the nucleus had thrown off an entire tail and was forming a new one.' Although several slit spectroscopic observations were attempted by Evershed, most useful observations were obtained with the prismatic camera set-up, similar to that employed for comet Daniel 1907d. The nucleus showed almost reflected solar spectrum, whereas the immediate surroundings of the nucleus (coma) showed strong bands of CN molecule at $\lambda\lambda$ 3871, 3883 and 4050 band (later identified as $C_3$) may be $C_2$ Swan bands. The monochromatic images of the tail showed bands at wavelengths similar to that observed in comet Daniel, i.e. the comet-tail bands due to $CO^+$ (Figure 16). The next appearance of Halley's comet occurred in 1986, which has been observed and studied in great detail at Vainu Bappu Observatory, Kavalur[52]. This study also discussed comparisons with observations of Michie Smith and Evershed.

'On May 19th 1910 the nucleus was computed to be in transit over the Sun. We attempted to photograph this event with the spectroheliograph, using the ultra-violet light of cyanogen which is a specially prominent radiation of the head of comets, but no trace of the comet could be seen in the photograph, nor could it be seen on the Sun in ordinary telescope. At this time the tail must have swept over the Earth and a sensation was caused in the newspapers by suggestion of a poisonous gas entering our atmosphere.'[2]

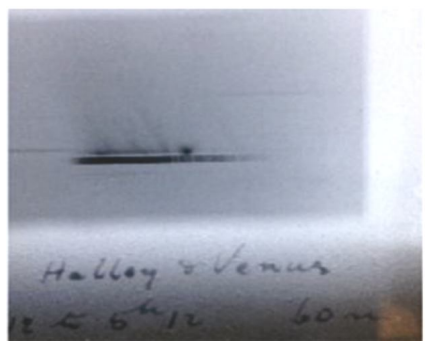

**Figure 16.** Prismatic spectrum of Halley's comet observed at Kodaikanal on 2 May 1910 showing the strong bands of $C_2$ and CN in and around the nucleus (coma). The monochromatic images of the tail show the comet-tail bands of $CO^+$. The spectrum immediately lower to the cometary spectrum is that of Venus. The spectrum extends from $H\beta$ to below $\lambda$ 3570.

Michie Smith retired in January of 1911 as Director of Kodaikanal Observatory and continued to live in Kodaikanal (Figures 17 and 18). His last scientific contributions are the observations of Halley's comet along with Evershed and list of prominences observed in the first half of 1910. Although he did not participate in the scientific activities at the observatory, nevertheless being in Kodaikanal he must have been following with interest the activities at the observatory till his death.

*New technique for measuring line shifts*

The large part of the spectroscopic work at Kodaikanal was devoted to the measurements of the minute shifts of the solar lines. Evershed[2] developed a technique of measuring accurate line shifts (radial velocities) by matching superposed spectra of negative and positive copies of the same spectrum. He even got a Hilger comparator built to measure such line shifts. 'I claim, however, for the positive on negative method that it extends our range of perception very appreciably.'[53] This method formed the basis of spectral comparators in use at several observatories. 'Its direct application to stellar spectra has been suggested by number of authors (particularly Babcock[54] and Fellgett[55]).'[56] Actually this principle also became the basis for the photoelectric radial velocity spectrometer of Griffin[56] and others.

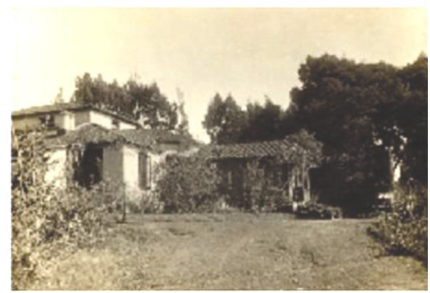

**Figure 17.** The residence of Charles Michie Smith after retirement in Kodaikanal.

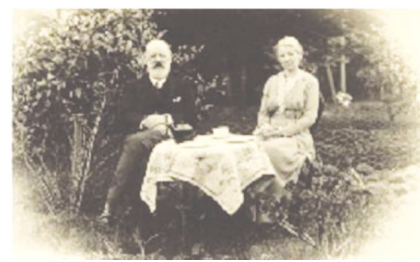

**Figure 18.** Charles Michie Smith with sister Lucy in Kodaikanal.

The minute shifts of the solar lines were found to be towards the red, especially the iron lines, when compared with spectra obtained with the electric arc. The shifts measured of the Sun's disk were found to agree with the gravitational redshift predicted from Einstein's general relativity, but at Sun's limb there seemed to be an extra redshift which could not be explained. Venus was used as a probe to measure redshift from the reflected sunlight spectrum by Evershed and Royds, which gave the same value as the disk.

*Second nova discovered at Madras*

A nova (new star) discovered G. N. Bower at Madras on 8 June 1918, between 3:30 and 4:30 pm G.M.T. in the constellation of Aquila created lot of excitement and hectic activity at Kodaikanal Observatory. Evershed came to know about it through a telegram from one C. L. Dundas, I.C.S. He also informed Evershed that it was equal magnitude to Altair ($\alpha$ Aquila) and the spectrum he obtained on June 10/11 when the star was of 0.2 magnitude (according to the estimate of Mary Ackworth Orr (Evershed's wife)[57] 'in addition to the bright hydrogen lines, he noted absorption lines in series, sharp towards the violet'. This is not the first nova that was discovered at Madras. Earlier Pogson discovered the recurrent Nova U Scorpii in 1863 at Madras Observatory. Evershed, in a hurry built a spectrograph and got a spectrum on 10 June 6:30 G.M.T., with the 15-inch Hyderabad lens[58] (Nizamia observatory's (note 23) and received help from its Director R. J. Pocock, who was visiting Kodaikanal at that time) in conjunction with an 18-inch siderostat and a prism spectrograph; 'it is of interest in showing no defined bright lines or bands but only a continuous spectrum with faint groups of absorption lines'[57].

This is the now famous nova designated as V603 Aquila (= nova Aquila 1918 III) which is the prototype of 'Fast novae'[59], similar to nova Persius 1901 (= GK Per) observed earlier by Naegamvala[26]. Systematic spectroscopic observations started at Kodaikanal with a new reconstructed prismatic camera using 'two excellent 45° prisms acquired from the dismantled Poona Observatory' and a 6-inch aperture photo-visual lens providing a spectrum of 104 mm length between H$\beta$ and H$E$ and at the red end also H$\alpha$. The observations continued from 10 June to 15 July, while the star's brightness changed from 0.2 to 4th magnitude. 'The hydrogen emission lines assumed extraordinary forms, H$\alpha$ appeared as a block of brilliant red light, the $\beta$ and $\gamma$ lines were extended blocks of blue and violet, well defined at their edges.' His observations revealed an absorption line system indicating expanding gas at 1600 and 2200 km/s that gets accelerated with time. He also identified the presence of nebular lines of $\lambda\lambda$ 5007, 4959 by 10 July, emergence of the so-called 'Orion Spectrum'[59]. He also could identify the lines of Fe II, Mg II in the expanding absorption system may be as shells of gas.

The nova has received lot of attention both spectroscopically[59,60] and photometrically[61] owing to its substantial outburst in which it brightened by 13 magnitudes. It was shown by Barnard[62] that the expanding ejecta had increased in diameter from 0.6 arcsec in October to 1.9 arcsec in December 1918 and reached a diameter of 26.0 arcsec by May 1933 (ref. 60). The ejecta has found to have had uniform expansion and was modelled by Weaver[63]. The nebular ejecta is now dispersed and no longer discernible. At maximum its brightness was –1.1 magnitudes. It was shown to be a cataclysmic binary with a period of $3^h19.^m5$ (ref. 64), consisting of a white dwarf in the system. Old novae like Nova Aquila 1918 provide good examples for the study of accretion phenomenon.

The high-resolution spectrographs built by Evershed were also used to study stellar physics problems like the line widths in stars like Sirius[65]. The broadening of spectral lines (shapes) due to various phenomena like Stark broadening, turbulence, rotation, etc. could have been an area where significant contributions could have been made at Kodaikanal, had such studies been continued be utilizing the 20-inch Bhavanagar telescope of Maharaja Takhatsingji's Observatory, which arrived (in boxes) in Kodaikanal in 1912. Although Evershed used the spectroscopic components like prisms, etc. of that telescope, the telescope itself was untouched. Of course, night-time astronomy was not a priority at Kodaikanal officially. About 40 years later, Das[43] in his report says 'A 20-inch Grubb reflector is being installed, and in the near future stellar work will form a regular feature of the activities of Kodaikanal'. Even this took another decade or two and another Director (M. K. Vainu Bappu) to get the stellar physics started in a regular way at Kodaikanal Observatory (and VBO at Kavalur).

Although Nizamiah Observatory (Osmania University), which got started in 1901, did mainly stellar astronomy by being one of the participants in that Carte-du-Ceil programme, physical astronomy was not practised until 1960s.

John Evershed retired by 1922 soon after Michie Smith's death and an era in the development of physical astronomy (observational astrophysics) came to an end.

**Michie Smith – promoter of astronomy at other centres**

Some of the responsibilities (or obligations) Michie Smith had as the Government Astronomer, Madras was to participate and advise other astronomical centres. One such institution was G. V. Jugga Row Observatory at Visakhapatnam[66]. Mrs A. V. Narasing Row, daughter of the founder and in-charge of the observatory, handed over the management of the Observatory to the Government of India on 8 November 1894 with an endowment fund of three lakh rupees for the permanent maintenance of the institution. The management was to be handled by a committee comprising the Collector (for the time being), meteorological reporters of the Government of Bengal and Madras, the Government astronomer and others. Michie Smith was thus included. John Eliot was also involved. Michie Smith took the job of ordering a camera and chronometers for the observatory, whose main aim was to photograph the sky. Choosing of the photographic plates and other needs were handled by the famous Oxford astronomer, H. H. Turner.

**Trees, bees and other things**

Michie Smith not only established the observatory, its buildings, instruments, etc., but also took care to plant trees which would provide proper environment for astronomical observations.

It was Michie Smith's inspired initiative that resulted in the planting of pine and oak trees in the observatory surrounding the domes and observing facilities that were helpful to improve the steadiness of the images. 'A large number of young trees have, however, been raised from seed and planted out whenever the weather is suitable. These young trees were largely pines of various kinds from the hills of Southern California for which the Director (CMS) is indebted to Mr Lukins of Pasedena. A number of seeds brought from the Lick Observatory were also grown and the seedlings promise to do well' (Annual Report of the Observatory, 1904). A part near the spectroheliograph buildings has been cleared and planted with good grass.

John Evershed, in a letter addressed to the then Director General of Observatories written in 1912, comments regarding the trees in the observatory grounds: 'These trees are mostly Pines planted by my predecessor Mr Michie Smith at his private expense, and in addition to their ornamental use are of great value in protecting the soil and rock surfaces from excessive heating by the sun. In this way they tend to improve the solar observations by preventing atmospheric disturbances near the instruments, and it follows that the Director incharge of the observatory will always endeavour to promote the growth of these trees and preserve as far as possible from destruction.' The same letter continues about other trees (wattles, blue gum, etc.). 'In the assistant Director's compound the land was not cleared when taken over from the Forest Department and the trees were purchased by Mr Michie Smith from the forest Department privately... they proved very useful for providing temporary buildings in the observatory and tripods, etc. for the erection of heavy instruments ... part of the equipment of Poona observatory when transferred to Kodaikanal was erected by the observatory staff at practically no cost to Government because timber was at hand for the work.'

Evershed' wife was deeply interested in the history of astronomy. She wrote several articles on the origin of constellations[67], and also completed her book *Dante and the Early Astronomers* in 1914 at Kodaikanal with help from her husband. She was also interested in the observations of prominences. Both husband and wife together published a long paper about their forms, distribution, connection with spots and other details (Figure 19) based on observations collected at Kenley, in England as well as at Kodaikanal[68].

## Horse named Jerusalem and Halley's comet

Michie Smith apparently maintained his own horse for moving in Kodaikanal (maybe elsewhere too), particularly between the town and the observatory. He named his horse Jerusalem. For many British officers of that time, horsemanship was a requirement. Even the foundation stone for Kodaikanal Observatory was laid by then Governor Wenlock after going there along bridle-path on a horseback in 1895. Being the president of the English Club (Kodaikanal Club) (note 24) Michie Smith used to spend time there. Apparently 'on one night in 1910 (April 1910) Mrs Peachey and her husband, the late Canon Peachey, were walking home in the evening from Tinnevelly settlement, they saw a comet and stopped at the English Club to ask Mr Michie Smith about it. He came out to look at it, said it must be Halley's comet, and immediately jumped on his horse, whose name was Jerusalem, and galloped up to the observatory to consult Mr Evershed[69].' (note 25). Not only did Michie Smith and Evershed make detailed observations and study of the Halley's comet, it apparently made quite an impression on Kodaikanal public as well. A letter written from Kodaikanal on 16 May 1910 is quoted as saying 'I was on Coaker's walk 4:40 this morning, and a more magnificent sight than the comet you can not imagine. It extended from horizon to mid-heaven'[69].

## Last days

In 1910, Michie Smith's services to the cause of science (or British raj) were recognized and he was awarded the title of 'Companion of the Indian Empire'. In the following year, he retired and settled in Kodaikanal. For number of years he was a member of the Municipality and even became its chairman for some years. He was apparently well known in Kodaikanal as the energetic and careful secretary and three-time president of the Boat Club, devoted to the protection of its property and rules. He also took interest in planting trees in the town and other places.

On one of his visits to home at Scotland in 1909, Michie Smith invited his younger sister Lucy to come and live with him in Kodaikanal. Lucy agreed and joined him, as his housekeeper, sometime in 1911 after quitting her job as a nurse and a matron at St. Thomas Hospital in London. In the words of her sister Alice '...the prospect of a peaceful retirement in the Indian hills must have held considerable appeal for her. Charlie had a beautiful home and garden beside the Kodaikanal Observatory. There were domestic animals too in plenty for Lucy to care for and talk to, just as she had done in the early days at Keig'. Lucy must have enjoyed life in India greatly. In 1914 Michie Smith accompanied by his sister Lucy, visited Australia just before the outbreak of war, to be present at the meeting of the British Association (Astronomers). In the spring of 1919, in spite of failing health, he and Lucy paid a visit to Scotland with a view to settle down there. But 'Charlie could no longer stand the harsh climate nor so he declared, the narrow Scottish mentality'. Both returned back to Kodaikanal. Soon after his arrival Charlie's physical and mental health started to deteriorate. Nursing him Lucy became sick and passed away six months before Charlie's death at the end of September 1922 (Figure 20). He bequeathed his 'considerable fortune' to his sister Bella, who had an 'unexpected windfall'[31]. 'All those who knew him well and appreciated his genuine kindly nature feel that in him they have lost a real friend (from *Madras Christian College Magazine* 1922)' (Figure 21).

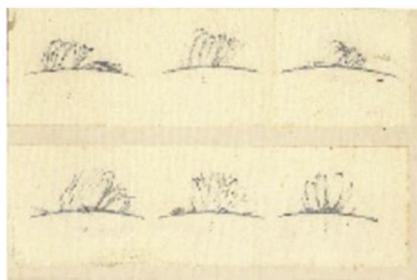

**Figure 19.** Solar prominences as sketched by John/Mary Evershed.

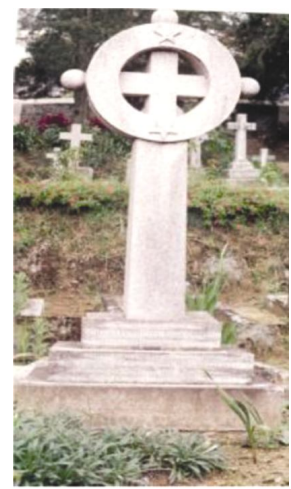

**Figure 20.** The memorial of Charles Michie and Lucy Smith in Kodaikanal.

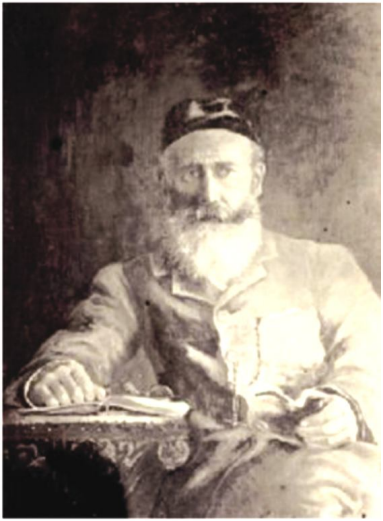

**Figure 21.** Painting of Charlie Michie Smith.

### Epilogue

Kodaikanal observatory is still an ongoing research facility living its 'raison d'être', whereas other solar (stellar) physics centres like Calcutta Observatory, Takhatasinghji Observatory or even Lucknow Observatory disappeared in much shorter time. They all were one-time wonders, that disappeared as soon as the main motivator (who started it) was no longer active or disappeared from the scene. The continuity was never kept up or the successors had very limited interest and diverted to other areas. The success of Kodaikanal was mainly because of its continuing motivated Directors and astronomers, till recently, who improved the facilities and provided innovative instrumentation, mainly people like John Evershed. They provided continuity of scientific purpose to the facility. Even after the British (trained) astronomers left the scene (Evershed, Royds), a smooth continuity prevailed and scientific activities continued with the same vigour by their successors (Narayana, Das and Bappu). New instrumentation was brought in to suit and pursue the newer and deeper problems.

It is often been said that the institutions like Kodaikanal remained as islands of British science. It may be relevant to ask how much of the science that was being done percolated to the natives (academic public) by these institutions? Have they been responsible to stimulate any research or interest in science among Indians? The reputation of the scientific activities at Kodaikanal led by Evershed did attract (inspire) few individuals like K. S. Krishnan and P. Chidambarum Aiyar (may be many more) for aspiring to work there. Krishnan got to know about Evershed and the solar physics work through his teacher Moffat at Madras Christian College. Unfortunately (or fortunately for C. V. Raman), he had to compete with Chidambarum Aiyar, who was already an assistant at Madras Observatory, thus an internal candidate, to the same position (second assistant) at Kodaikanal Observatory. The authorities prevailed on Evershed to accept Chidambarum Aiyar[70].

Another puzzling aspect is, why was there no communication between institutions (or astronomers) pursuing the same science, e.g. Naegamvala with Madras or Kodaikanal astronomers. We are puzzled why M. N. Saha never interacted with Kodaikanal astronomers like Evershed or Royds when he was developing and applying his theory of ionization to stellar spectra while complaining about lack of access to observational data, particularly about solar and sunspot spectra. Although he was communicating with people like Fowler, Milne in England and H. N. Russell at Princeton, USA, Saha did not appear to have made an attempt to get in touch with Evershed or Royds, who probably would have provided him with observational data. Apparently Gilbert Walker, then Director General of Observatories, had invited Saha to join the Kodaikanal Observatory as assistant Director to Evershed. 'Though Saha declined, it did not diminish Walker's support'[71] to Saha. About Saha's refusal to accept Kodaikanal offer, Anderson[72] writes 'the reason for the refusal in the authorized biography is hardly convincing'. He then quotes Sen 'Working under a man of Evershed's genius and genial personality would have been a great honour and privilege, and it probably would have brought him (Saha) greater recognition, but his heart was in physics, and he would not change it even for astrophysics.' Anderson further comments 'This is curious (retrospective) statement by Saha because Saha's reputation was already established in astrophysics'.

Saha's contact with Kodaikanal Observatory occurred much later when he visited as chairman of post-war committee for development of astronomy and astrophysics in India in 1945–46. The committee had Krishnan also as a member, although he never visited Kodaikanal or participated in its meetings[70].

### Notes

1. 'I can remember also, at the age of six, seeing in an illustrated paper a picture of German shells falling in the streets of Paris, and it was during this siege by the Germans that Janssen, the French astronomer, escaped by balloon from the besieged city to observe a total eclipse of the Sun[2].' It was 1870 eclipse in which the totality passed through Spain, Sicily, North Africa. Janssen escaped from Paris by a balloon and went to Algeria to observe the event, even though Norman Lockyer arranged for free passage to him through German lines[3].
2. It is now known[6,7] as the green line of 5303 Å of Fe XIV, initially thought to be due to an unknown element designated as coronium.
3. Apparently in Tamil, Kodaikanal means 'summer, a mirage', implying hardly any summer days[9].
4. The quote from Pogson says $\mu$ Argus. The constellation Argus was divided later into three constellations, Pupis, Carina and Velorum. $\mu$ Argus refers to $\mu$ Velorum. But $\mu$ Velorum is a solar-type star and does not match the description of the spectrum presented. Subsequent check revealed, from the report of J. Eliot, who went through the unpublished records of Pogson after his death, that the unpublished observations are of $\gamma$ Argus ($\gamma$ Velorum). $\gamma$ Velorum is now a well-studied Wolf-Rayet star binary[11].
5. Refers to T CrB, the cataclysmic variable–recurrent nova which had an outburst in 1866 and showed emission lines of hydrogen and helium.

6. For biographical details and his pay at various times, see Kochhar and Narlikar[19].
7. 5007 Å line; attributed to an unknown element called nebulium, later identified as due to [O iii] by Ira S. Bowen in 1920s.
8. This and other letters of Naegamvala mentioned in the text were obtained from RGO collection.
9. See Glass[24] and the references therein for the history. This is also the most travelled telescope (for its size) in the country. Starting at Poona it travelled to Kodaikanal in 1912, where it was in boxes till 1951, then to Kavalur and later to Leh and back again to Kodaikanal[25].
10. See Anupama and Kantharia[27] for recent study and image of the remnant of this nova obtained with the 2-m Himalayan Chandra telescope of IIA.
11. These and other official correspondences are obtained from Archives of Tamil Nadu.
12. Kashi Nandy managed to get some money from Royal Society, Edinburgh to support my visit to ROE in 1984 (N.K.R.).
13. This refers to [O I] 5577 Å line – likely due to airglow.
14. The largest telescope Kodaikanal ever seen, the 20-inch Bhavanagar telescope of Maharaja Takhatsingjii Observatory, arrived from Poona in 1912 after Naegamvala's retirement.
15. A more recent site survey at Leh was conducted in 1980–89 and similar conclusions were arrived at[35].
16. The Scottish connection of CMS. Lord Kelvin (William Thomson) being a personal friend of William Robertson Smith, the elder brother of Michie Smith, knew CMS well.
17. This estimate was provided by Eliot for the establishment of the hill observatory.
18. The other members of the committee present were General Walker, Sir G. G. Stokes, Dr Huggins, General Strachey, Astronomer Royal, while Dr Common could not attend. However, Michie Smith was in constant touch with Astronomer Royal.
19. Among other rituals being practised at the time of the eclipse like taking holy dips in the Ganges, etc. apparently the Nizam of Haidarabad has released 50 prisoners, each of whom received a present of money and clothes on the occasion.
20. See Pang[41] for a delightful account of her experiences in the field and at the eclipse camp at Jeur.
21. It was proposed that two out of the three domes were to be constructed and the third dome was to have been erected and left unoccupied in the hope that, in the future, a large telescope may be placed there which, of course, never materialized.
22. On 16 July 2013 the telegram service was discontinued for good by the Indian Postal Department.
23. See Bhaskaran[58] for the history of the telescope and loaning of the lens.
24. Apparently it had Indian, American, Scottish, Welsh and Irish members as well, although it is called 'English'
25. Does not say much about anticipating the arrival of history's most famous comet.

ACKNOWLEDGEMENTS. We thank Ms Astrid Hess, grand niece of Michie Smith (CMS), for generously providing us with several photographs and information on CMS, his family and early days of Kodaikanal Observatory. Dr Joshua Kalapati of Madras Christian College (MCC), Chennai provided the photographs and CMS's obituary report in MCC magazine, 1922. We also thank our IIA library colleagues (B. S. Mohan, P. Prabahar, P. N. Prabhakara and M. Venkatesha) for help at various stages. N.K.R. thanks Dave Pike for photographs of the CMS family memorial, which mentions Michie Smith at the bottom and the tombstone of Robert William Smith at Keig, Aberdeenshire. We also thank Dipankar Mallik for useful comments on an earlier version of the paper.

*N. Kameswara Rao\*, A. Vagiswari and Christina Birdie are in the Indian Institute of Astrophysics; Bangalore 560 034 India; N. Kameswara Rao is also at The W.J. McDonald Observatory, University of Texas, Austin, TX78712-1083, USA.*
*\*e-mail: nkrao@iiap.res.in*